\documentclass{aa}

\usepackage{amsmath}
\usepackage{multirow}
\usepackage[T1]{fontenc}
\usepackage[normalem]{ulem}
\usepackage{hyperref}

\begin{document}

\title{Streaming instability in a global patch simulation of protoplanetary disks}

\author{Mario Flock\inst{1} \and Andrea Mignone\inst{2}}

\institute{Max-Planck-Institut f{\"u}r Astronomie, K{\"o}nigstuhl 17, D-69117 Heidelberg, Germany\\ \email{flock@mpia.de} \and Dipartimento di Fisica, Universitá di Torino via Pietro Giuria 1 (I-10125) Torino, Italy}

\date{Received December 10, 2020; accepted ...}

\abstract
{}
{In the recent years, sub/mm observations of protoplanetary disks have discovered an incredible diversity of substructures in the dust emission. 
An important result was the finding that dust grains of mm size are embedded in very thin dusty disks. This implies that the dust mass fraction { in the midplane} becomes comparable to the gas, increasing the importance of the interaction between the two components there.} 
{We address this problem by means of numerical {  2.5D} simulations in order to study the gas and dust interaction in fully global stratified disks. 
To this purpose, we employ the recently developed dust grain module in the PLUTO code. Our model focuses on a typical T Tauri disk model, simulating a short patch of the disk at 10 au which includes grains of constant Stokes number of $St=0.01$ and $St=0.1$, corresponding to grains with sizes of 0.9 cm and 0.9 mm, respectively, for the given disk model.}
{By injecting a constant pebble flux at the outer domain, the system reaches a quasi steady state of turbulence and dust concentrations driven by the streaming instability. 
For our given setup and using {  resolutions up to 2500 cells per scale height} we resolve the streaming instability, leading to local dust clumping and concentrations. 
Our results show dust density values of around 10-100 times the gas density with a steady state pebble flux between $3.5 \times 10^{-4}$ and  $2.5 \times 10^{-3} M_{\rm Earth}/\mathit{year}$ for the models with $\mathit{St}=0.01$ and $\mathit{St}=0.1$.} 
{The grain size and pebble flux for model $\mathit{St}=0.01$ compares well with dust evolution models of the first million years of disk evolution. For those grains the scatter opacity dominates the extinction coefficient at mm wavelengths. These types of global dust and gas simulations are a promising tool for studies of the gas and dust evolution at {  pressure bumps} in protoplanetary disks.}

\keywords{protoplanetary disks --- hydrodynamics --- streaming instability}

\maketitle
\section{Introduction \label{sec:intro}}
%
%
%

The interaction between the gas and the dust is a crucial step in the process of planet formation. Once grains collide in protoplanetary disk, they are able to stick and grow which leads them to decouple from the gas motion \citep{saf72,whi72,ada76,weidenschilling77,wet93}. 
Once grains settle to the midplane, they concentrate speeding up the dust coagulation and growth even further \citep{wei97,ste97,wei00,lai08,brauer:2008,birnstiel:2011}. 
Depending on the radial profiles of temperature and density, these dense dust layers are prone to other types of instabilities, especially the gas and dust drag instabilities \citep{squ18,hop18,zhu20}. 
The streaming instability (SI henceforth), one important subclass of these drag instabilities, has been studies extensively in the recent two decades, in particular for its role in explaining planetesimal formation \citep{youdin05,Johansen:2007,Bai:2010,Yang:2014,carrera:2015,yan17,Schreiber:2018,carrera:2020}. 
The growth rate of the SI depends on the local dust-gas-mass ratio and can reach values of around the orbital timescale for dust-to-gas mass ratios close to unity \citep{squ18,pan20}. 
Recent works focused on the SI with multi-grain species \citep{lai14,krapp:2019,zhu20b,par20} demonstrating that its growth rate is in general reduced when considering multiple grain sizes. 
Lagrangian and fluid methods were used in the past and both have their advantages and disadvantages. 
As Lagrangian methods introduce a fixed number of grains, in stratified disk models this means that they can only resolve a certain height of the dust disk { and one has to ensure for a good sampling to supress the noise level \citep{cad19}.} 
On the other hand, they allow to follow individual grain motions becoming particularly suited to study larger grains which decouple from the gas motion. 

Recent studies emphasized again the importance of the level of gas turbulence to determine where the SI can operate \citep{jaupart:2020,umurhan:2020}. 
So far most of the simulations have been performed in local box simulations, and only recently first simulations appeared using global unstratified simulations \citep{Kowalik2013,mig19} confirming the main characteristics of the SI. Very recently \citet{schaefer:2020} investigated the interplay between the vertical shear instability and the SI in stratified global models, demonstrating the importance of the large scale gas motions for the dust concentrations. 


In this work we propose a framework to study the SI in global stratified disk simulations with more realistic conditions of the radial pressure gradient profile and the pebble flux. 
For the first time to the extent of our knowledge, the SI is investigated in a high resolution stratified global simulations using spherical geometry. 
The challenge is twofold: first, the SI requires resolutions of several hundreds cells per gas scale height and second, the finite extent in the radial direction introduces a time limit to study the SI as the grains radially drift through the domain. 
In Section 2 we explain the numerical method and the disk setup, while in section 3 we present the results and compare them to local box simulation, emphasizing the role of the pebble flux compared to the classical total dust-to-gas mass ratio. 
Finally we test our results with constraints from typical T Tauri star disk systems and dust evolution models and give estimates on the optical depth at mm wavelengths. We present the discussion and conclusion in section 4 and 5.

\section{Methods and disk setup}
%
%

In order to setup our model we follow the work of \citet{nak86} which prescribes the dust and gas velocities ($v$ and $V$, respectively) using cylindrical geometry $(R,\phi,Z)$ as
\begin{align}
    V_R &=& -\frac{\rho}{\rho+\rho_{\rm d}} \frac{2 D \Omega_{\rm K}}{D^2+\Omega_{\rm K}^2}\eta R \Omega_{\rm K} \label{eq:vrd}\\
    V_\phi &=& \left( 1-\frac{\rho}{\rho+\rho_{\rm d}} \frac{D^2}{D^2+\Omega_{\rm K}^2}\eta \right ) R \Omega_{\rm K} \label{eq:vpd}\\
    v_R &=& \frac{\rho_{\rm d}}{\rho+\rho_{\rm d}} \frac{2 D \Omega_{\rm K}}{D^2+\Omega_{\rm K}^2}\eta R \Omega_{\rm K} \label{eq:vrg}\\ 
    v_\phi &=& \left [ 1+\left ( \frac{\rho_{\rm d}}{\rho+\rho_{\rm d}} \frac{D^2}{D^2+\Omega_{\rm K}^2}-1\right)\eta \right ] R \Omega_{\rm K}, \label{eq:vpg}
\end{align}
where  $\rho$ and $\rho_{\rm d}$ denote, respectively, the gas and dust density, $\Omega_{\rm K}=\sqrt{GM/R^3}$ is the Keplerian frequency with the gravity constant $G$, $M$ is the mass of the star. 
We define the factor
\begin{equation}
  D=\frac{1+ \epsilon }{St}\Omega_{\rm K}    
\end{equation}
with the dust to gas mass ratio $\epsilon=\rho_{\rm d} / \rho$ and the dimensionless Stokes number $ \mathit{St}=t_s \Omega_{\rm K}$ with $t_s$ being the stopping time.
Likewise, we introduce
\begin{equation}
  \eta=- \frac{1}{2 \rho R \Omega_{\rm K}}\frac{\partial P}{\partial R}\Omega_{\rm K}    
\end{equation}
where $P$ is the (gas) pressure. 
We note here the factor of $1/2$ which was also adopted by \citet{you07}. 
With this, the pure gas azimuthal velocity translates to $v_\phi=(1-\eta)R \Omega.$\footnote{We note that some investigators define $\eta$ without the factor $1/2$, see also \citet{takeuchi02}.} 
We assume a local isothermal equation of state with the pressure defined by $P=c_s^2 \rho$ with $c_s(R)$ being the speed of sound. 
%
%
The scale height $H$ of the gas is defined as $H=c_s/\Omega_{\rm K}$
with radial dependence
\begin{equation}
    H=H_0 \left ( \frac{R}{R_0}\right )^\frac{Q+3}{2},
\end{equation}
where $P$ and $Q$ are the radial profile exponents for the density and temperature $T \sim c_s^2$.

The initial profile of the gas density in the $R-Z$ plane is set by
\begin{equation}\label{eq:gas}
    \rho (R,Z)= \rho_0 \left ( \frac{R}{R_0} \right )^P \exp\left[\frac{R^2}{H^2} 
                  \left( \frac{R}{r} - 1 \right)\right]\,,
\end{equation}
where $r = \sqrt{R^2 + z^2}$ is the spherical radius.

The dust density $\rho_{\rm d}$ is defined similarly to Eq. (\ref{eq:gas}) with the dust scale height $H_{\rm d}$: 

\begin{equation}\label{eq:dust}
    \rho_{\rm d} (R,Z)= \rho_{d,0} \left ( \frac{R}{R_0} \right )^P 
    \exp\left[\frac{R^2}{H_{\rm d}^2} \left( \frac{R}{r} - 1\right)\right]
\end{equation}
Finally, the initial total dust to gas mass ratio is defined through the ratio of the vertically integrated surface densities, that is,  $\Sigma_{d,0}/\Sigma_0$, related to the midplane densities as
\begin{equation}
    \rho_0   = \frac{\Sigma_0}{\sqrt{2 \pi} H_0}\,,\qquad
    \rho_{d,0} = \frac{\Sigma_{d.0}}{\sqrt{2 \pi} H_{d,0}},
\end{equation}
for the gas and for the dust using $\Sigma_{d,0}$ and $H_{d,0}$ respectively. The parameter for the model are summarized in Table~\ref{tab:init}.

\subsection{Numerical configuration}

Our computations are performed in spherical geometry using the HLL Riemann solver and the $2^{\rm nd}$-order Runge Kutta integration in time to advanced conserved variables.
Piece-wise linear reconstruction with the MC limiter has been used.
The gravity force is handled by adding the vector force on the right hand side of the momentum equation for both gas and particles. 
The radial boundary conditions for the hydro variables are zero-gradient without allowing for material to enter the domain. { This is obtained by setting the normal velocity component to zero in case the velocity in the active domain is pointing inward.}
At the meridional boundary, ghost zones are filled by extrapolating the exponential profile of the gas density while the remaining variables are set to have zero-gradient. 
Similarly to the radial boundaries, gas is not allowed to enter the domain. 
Dust particles are advanced in time using the exponential midpoint method and the cloud-in-cell (CIC) weighting scheme is applied to determine the gas values at the grain position \citep{mig19}. 
Gas and dust are coupled by mutual feedback terms accounting for drag force effects. 
As particles are stored locally on each processor, we reach best parallel performance when we use a decomposition $X:2$ for the $r:\theta$ domain. 
For more information about the method we refer to our previous work \citep{mig19}.

We note that the resolution was chosen to resolve the grains drag and therefore the grid size should fulfill
\begin{equation}
    \Delta x < t_s c_s 
\end{equation} 
to resolve the grains stopping length. 
For our setup with a fixed Stokes number this becomes $\Delta x < St H$. Further we have to resolve the SI for the regimes of $\mathit{St}=0.1$ and $\mathit{St}=0.01$ and we adopt a grid resolutions similar to that used by \citet{yan17} with around $\sim 1000$ cells per $H$. The aspect of resolution is discussed in detail in  Section~\ref{sec:res}. { We use a uniform grid spacing in the radial and in the $\theta$ direction. The total number of grid cells and domain extent are given in Table~\ref{tab:model}.}

 \begin{table}
\begin{tabular}{lll}
  \hline
  \hline
$St$    & $10^{-2}$ and $0.1$\\
$H_0/R$     & 0.07\\ 
$H_{d,0}/R$   & 0.0014\\
$\Sigma_0$ & 60 $g\, cm^{-2}$\\
$\Sigma_{d,0}$ & 0.6 $g\, cm^{-2}$\\
$R_0$        & 10 au\\
P            &  -1\\
Q            &  -1\\
  \hline
  \hline
\end{tabular}
\caption{Initial setup parameters for the 2D dust and gas disk models.}
\label{tab:init}
\end{table}

 \begin{table}
\begin{tabular}{llllll} 
Run & Domain  & Grid & $ \frac{H}{\Delta r}$ & $St$ & \#\\
\hline
\texttt{ST1} & $ \rm \pm 0.7:\pm \frac{7}{1000}$ &  1024x128  & 640 & $10^{-1}$ & 1.04$\times$$10^7$\\
\texttt{ST2} & $ \rm \pm 0.7:\pm \frac{7}{1000}$ &  4096x512  & 2560 & $10^{-2}$ & 1.04$\times$$10^7$\\

\end{tabular}
\caption{Starting from the left: model name, extent of the domain ($10 \rm au \pm \Delta R$au  : $\pm \Delta \theta $), total grid resolution, resolution per gas scale height, Stokes number and initial number of particles.}
\label{tab:model}
\end{table}

\subsection{Lagrangian particle setup}

In order to sample the dust density we introduce individual particles to the simulation. Those particles represent a swarm of grains.   
Similar to the work by \citet{yan17}, we start the simulation assuming a dust scale height which is reduced compared to the gas scale height. 
This increases locally the dust to gas mass ratio and enhances the growth rate of the SI. 

We present our results using particle sample runs with approximately 80 particle per cells for model \texttt{ST1} and 5 particles per cell for model \texttt{ST2}. For our models we determined that a sampling of around 5 cells or more is needed for consistent results. More details on the sampling and a benchmark can be found in appendix A and appendix B.

The total dust mass in the domain is 
\begin{equation}
M_{\rm d}=\int_r \int_\phi \Sigma_{\rm d}(r) r \mathrm{d}r \mathrm{d}\phi = 2 \pi \Sigma_{\rm d,0} R_0 \Delta R \sim 1.978 M_{\rm Earth}.
\end{equation}
Using a total number of grains $N_{\rm tot}=1.04\times 10^7$, we sample the dust mass in our domain with particle swarms with masses of $1.902\times 10^{-7}$ Earth masses.

Once the mass of the particle is fixed, the number of particles in each cell can be determined with $N_{cell}= (\rho_{\rm d} \Delta V)/m_g$.
A vertical profile of the detailed sampling over height is shown in the appendix A.

\begin{figure*}[t]
    \resizebox{\hsize}{!}{\includegraphics{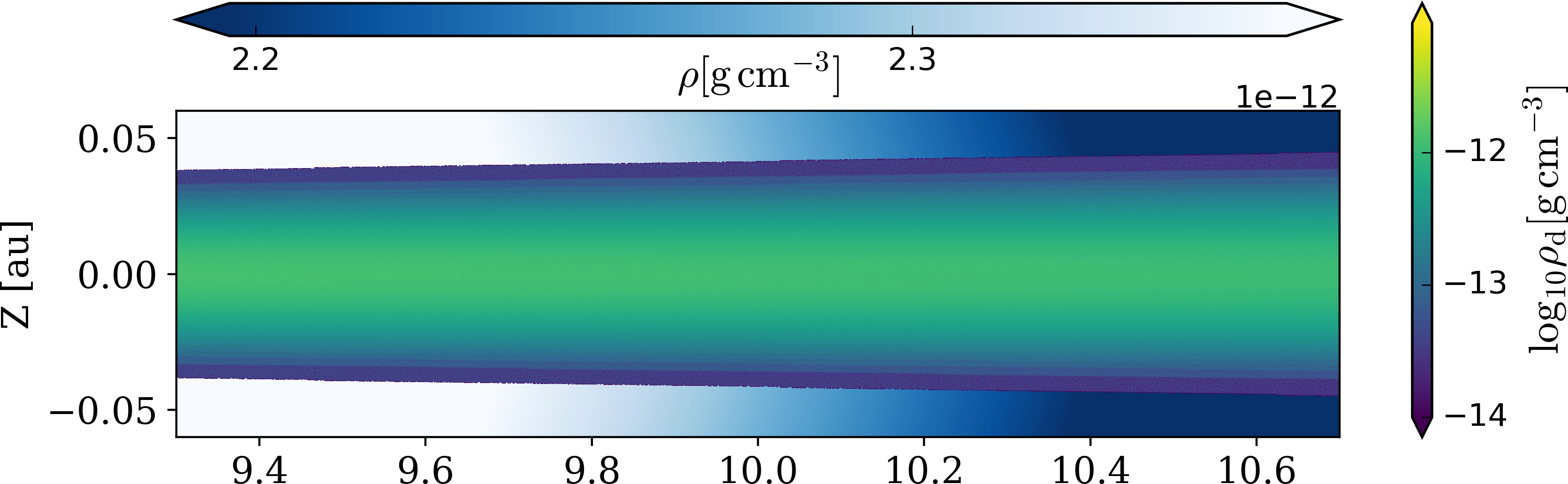}}
    \resizebox{\hsize}{!}{\includegraphics{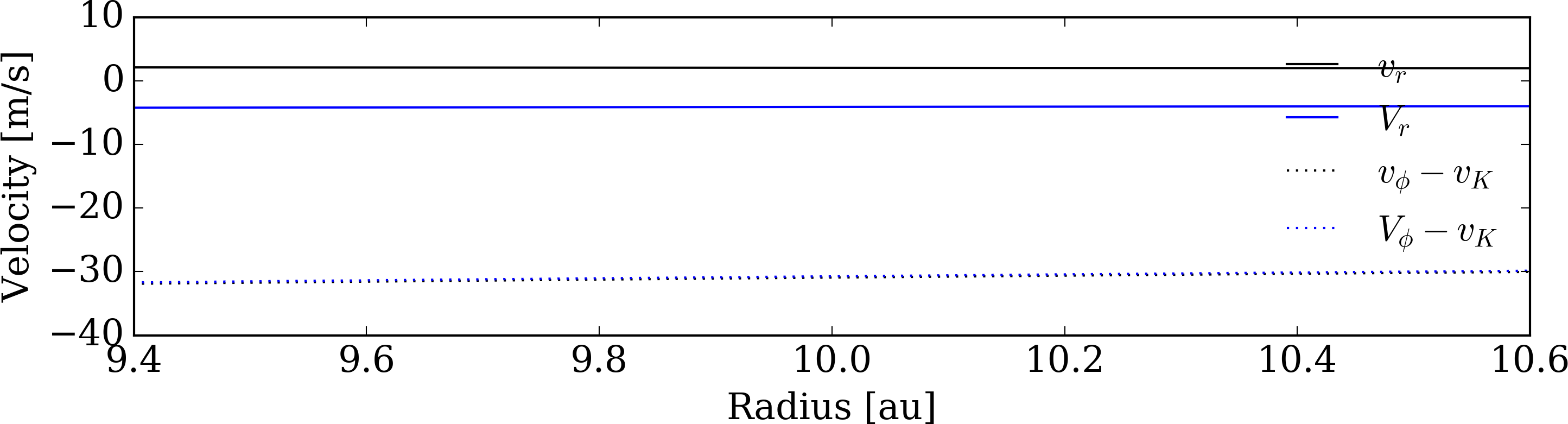}}
    \caption{Top: Initial distribution of the gas and dust density in the R-Z plane. Bottom: Initial radial profile of the gas and dust velocities at the midplane.}
    \label{fig:dg}
\end{figure*}

The initial profile of the gas and dust density in the R-Z plane is shown in Fig.~\ref{fig:dg}. 
As seen in the contour plot, the particle method can only resolve a certain vertical extent which depends on the total number of grains. 
The initial gas and dust velocities are plotted in the bottom panel of  Fig.~\ref{fig:dg}, following the equilibrium solutions, Eq. (\ref{eq:vrd}), (\ref{eq:vpd}), (\ref{eq:vrg}) and Eq. (\ref{eq:vpg}).

In this equilibrium solution at the midplane, the grains slowly drift radially inward while the gas slowly drifts radially outward.

\subsection{Damping zones}
{ To prevent numerical effects from the radial domain boundary we need to implement buffer zone in which the variables of density and velocity are relaxed back to their initial value. To implement those, we apply a wave killing zones close to the radial inner and outer boundary to reduce the interaction with the boundary. Similar buffer zones were already tested and implemented in our previous work \citet{mig19}.} 
In this zones, the variables are relaxed to the initial equilibrium values,
                                  
\begin{equation}
\label{eq:damp}
  Q(\pmb{\vec{x}},t) =   Q_0(\pmb{\vec{x}})
                 + \Big[Q( \pmb{\vec{x}},t) - Q_0(\pmb{\vec{x}})\Big]e^{F(R) \Delta t/T}\,,
\end{equation}
where $Q(\pmb{\vec{x}},t)$ represents either the radial or azimuthal fluid velocity, $Q_0(\vec{x})$ is the corresponding equilibrium value, $T = 1.0$ and
\begin{equation}
   F(R) = 2 - \tanh\left(\frac{R - R_b}{w}\right)^8
            - \tanh\left(\frac{R - R_e}{w}\right)^8
\end{equation}
is a tapering function with $w = 0.05$.

{ A second important point is to prevent the dust dragging the gas material out. As we constantly inject new grains at the radial outer zone we also have to replenish the gas material. To study the SI in a quasi steady state configuration, we implement a density relaxation which relaxes the loss of gas mass to the initial value. For each grid cell we apply  
\begin{equation}
  \rho(r,\theta) =   \rho_0(r,\theta)
                 + \Big[\rho((r,\theta),t) - \rho_0(r,\theta)\Big]e^{2 \Delta t/T}\,,
\end{equation}
setting the parameter $T=1.0$. In appendix C we show that the influence of parameter T in this regime of the density relaxation remains small.}

\subsection{Injection and destruction of particles}
At the outer boundary, new particles must be constantly injected as the radial drift empties these buffer zones. 
Refilling has to be done carefully as in this model the vertical settling quickly changes the structure in the buffer zones. 
To prevent that the solution is quickly shifted away from the equilibrium solution we damp the vertical velocity as long as the particles remain in the outer buffer zone. 
{More specifically,} at every timestep we set
\begin{equation}
  v_\theta(r,\theta) =   (1-\chi)v_\theta(r,\theta)
\end{equation}
in the radial outer buffer zone, using $\chi=10^{-5}$. { We found that this efficiently prevents the particle from settling already to the midplane before they have entered the active domain.} 
In addition, we inject only dust when the dust density drops below a certain factor $f$, which is $\rho_{d} < f \rho_{d,0}$ with $f=0.5$ for $St=0.1$ particles and $f=0.25$ for $St=0.01$. 
In this way, we are able to resupply the disk with a constant pebble flux without any accumulation at the buffer zone edges. 
Particles are constantly injected in $r\in[10.6,\,10.7]$ au  as they radially drift inward. 
For $r < 10.6$ au they start to settle eventually triggering the SI. 
At around 10.5 au and inwards, the structure and the dynamics of the dust and gas remains self-similar. 

Particles are removed from the computational domain once they cross the inner boundary at 9.3 au. To avoid dust accumulation at the inner buffer zone we reduce the azimuthal velocity through
\begin{equation}
  v_\phi(r,\theta) =   (1-\chi)v_\phi(r,\theta)
\end{equation}
at every timestep in the inner buffer zone using $\chi=10^{-8}$. 
This reduces their rotational velocity thus increasing their radial drift to avoid any concentration close to the buffer zone edge at 9.4 au.

\subsection{Particle size}
In our models we fix the Stokes number of the grains which makes a comparison to previous works easier. 
As the gas density in our domain does not vary much, we can determine approximately the dust grain sizes. 
For our disk setup we are in the Epstein regime and we can determine the size using 
\begin{equation}
a=\frac{St \rho H}{\rho_{\rm grain}}
\end{equation}
where, for the grain density, we employ $\rho_{grain}=2.7\rm \, g \, cm^{-3}$.
Using the midplane density at 10 au we determine the grain sizes of 0.88 mm for model \texttt{ST2} and 8.8 mm for model \texttt{ST1}.

\section{Results} 
%
%

%
%
 %


\begin{figure}
    \resizebox{\hsize}{!}{\includegraphics{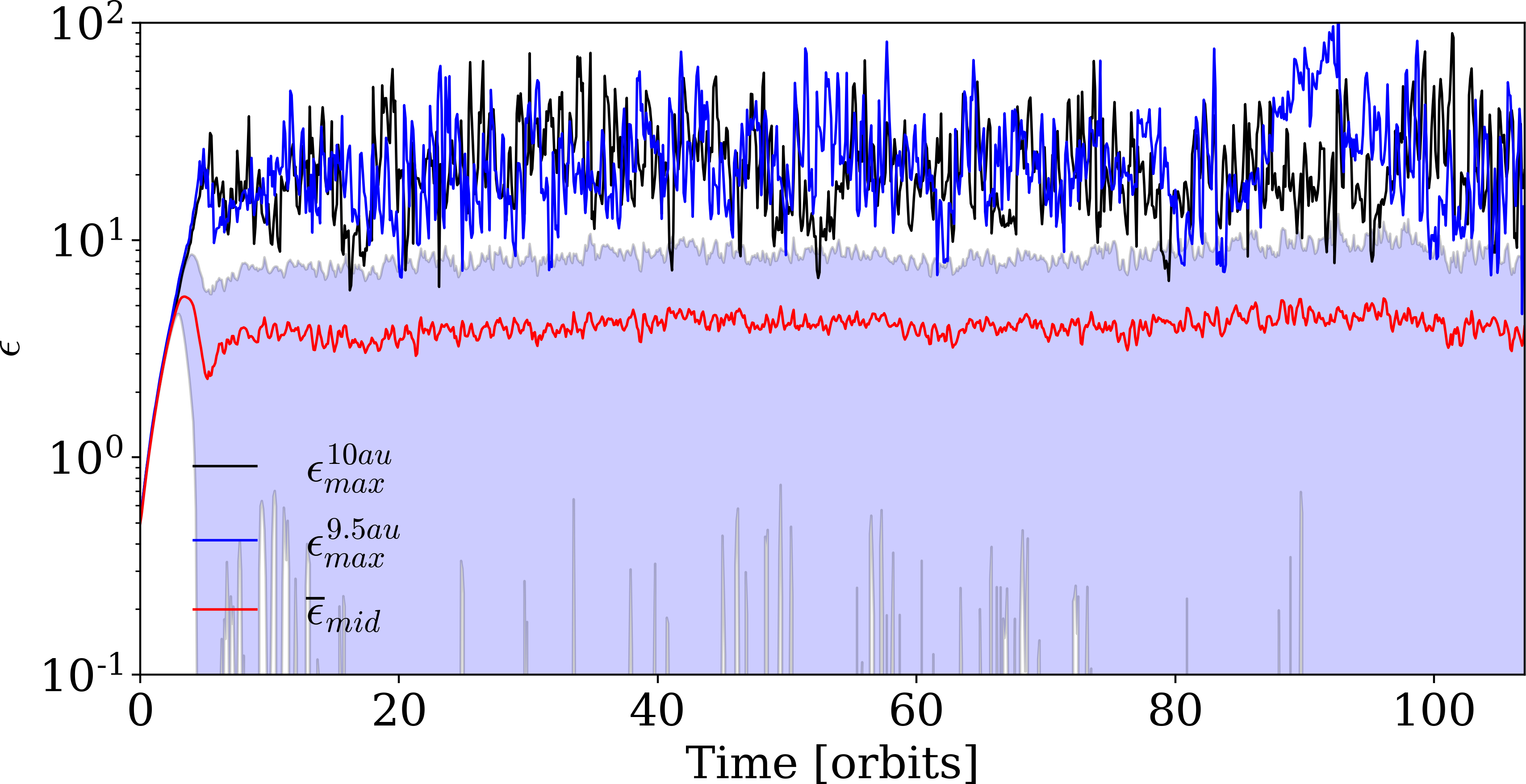}}
  \resizebox{\hsize}{!}{\includegraphics{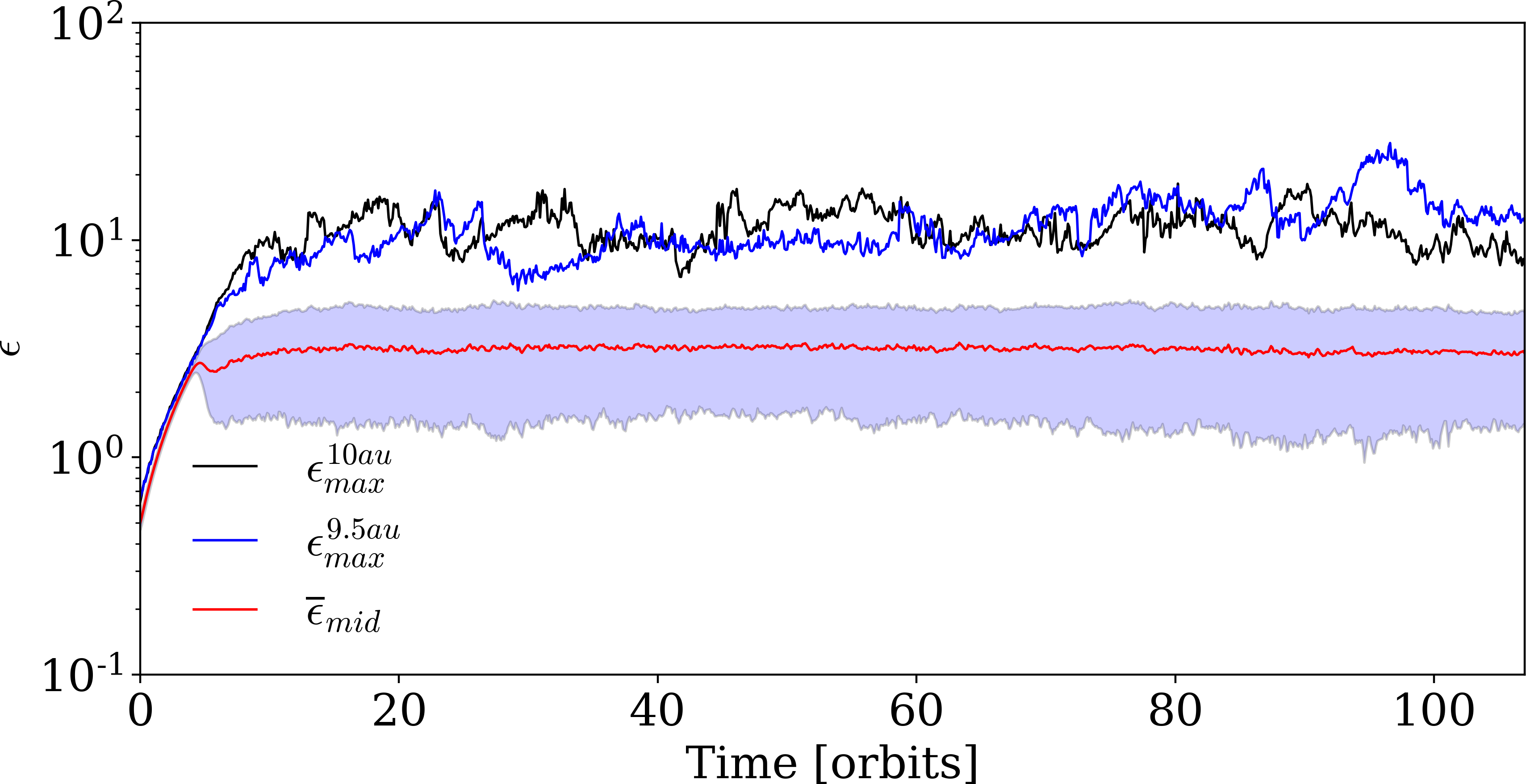}}
    \caption{Maximum dust to gas mass ratio at 9.5 au (blue) and at 10 au (black) for model \texttt{ST1} (top) and model \texttt{ST2} (bottom). The space averaged dust to gas mass ratio at the midplane is shown with the red solid line including the standard deviation (filled color).}
    \label{fig:dgrowth}
\end{figure}


\begin{figure}
    \resizebox{\hsize}{!}{\includegraphics{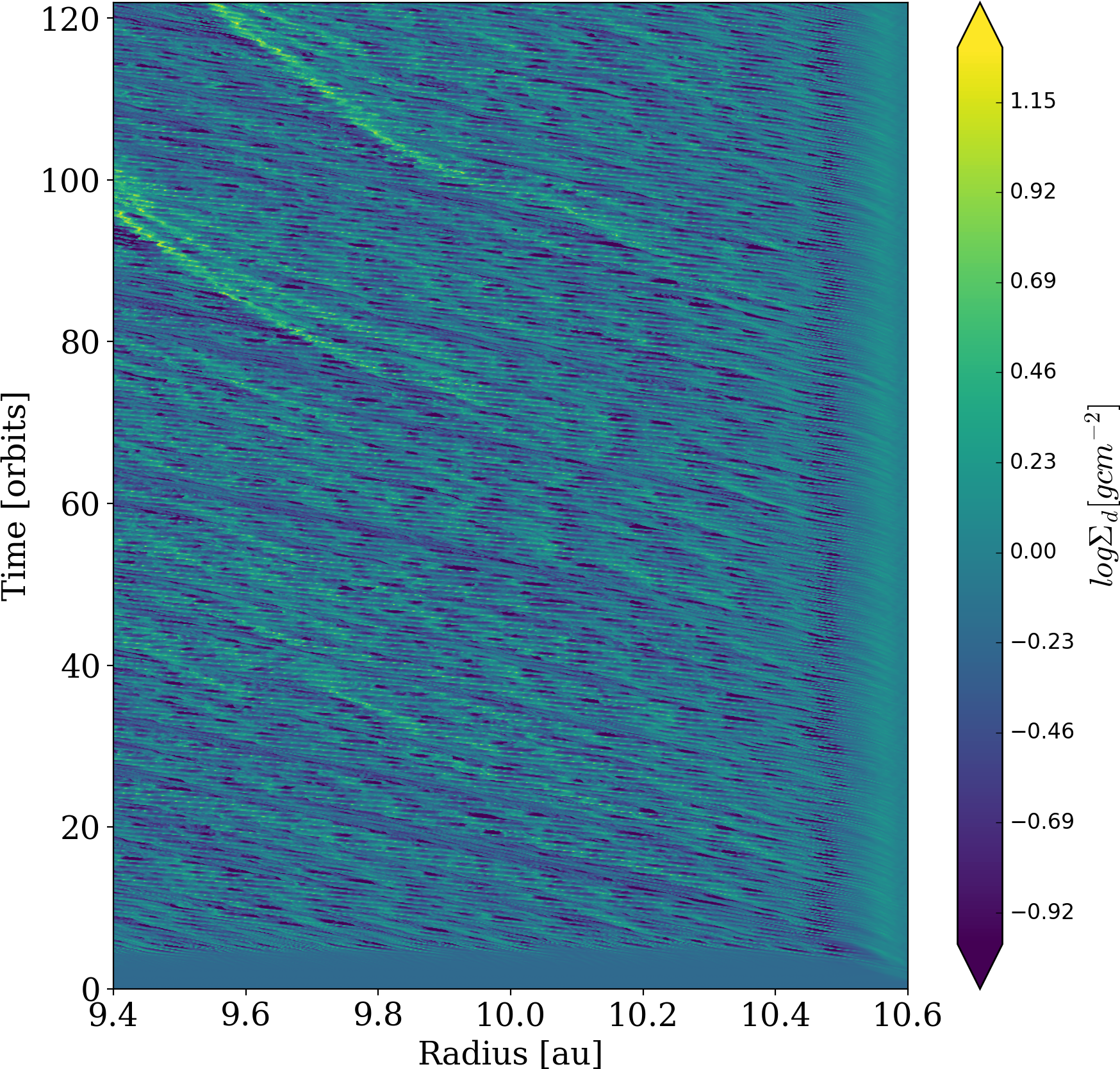}}
    \resizebox{\hsize}{!}{\includegraphics{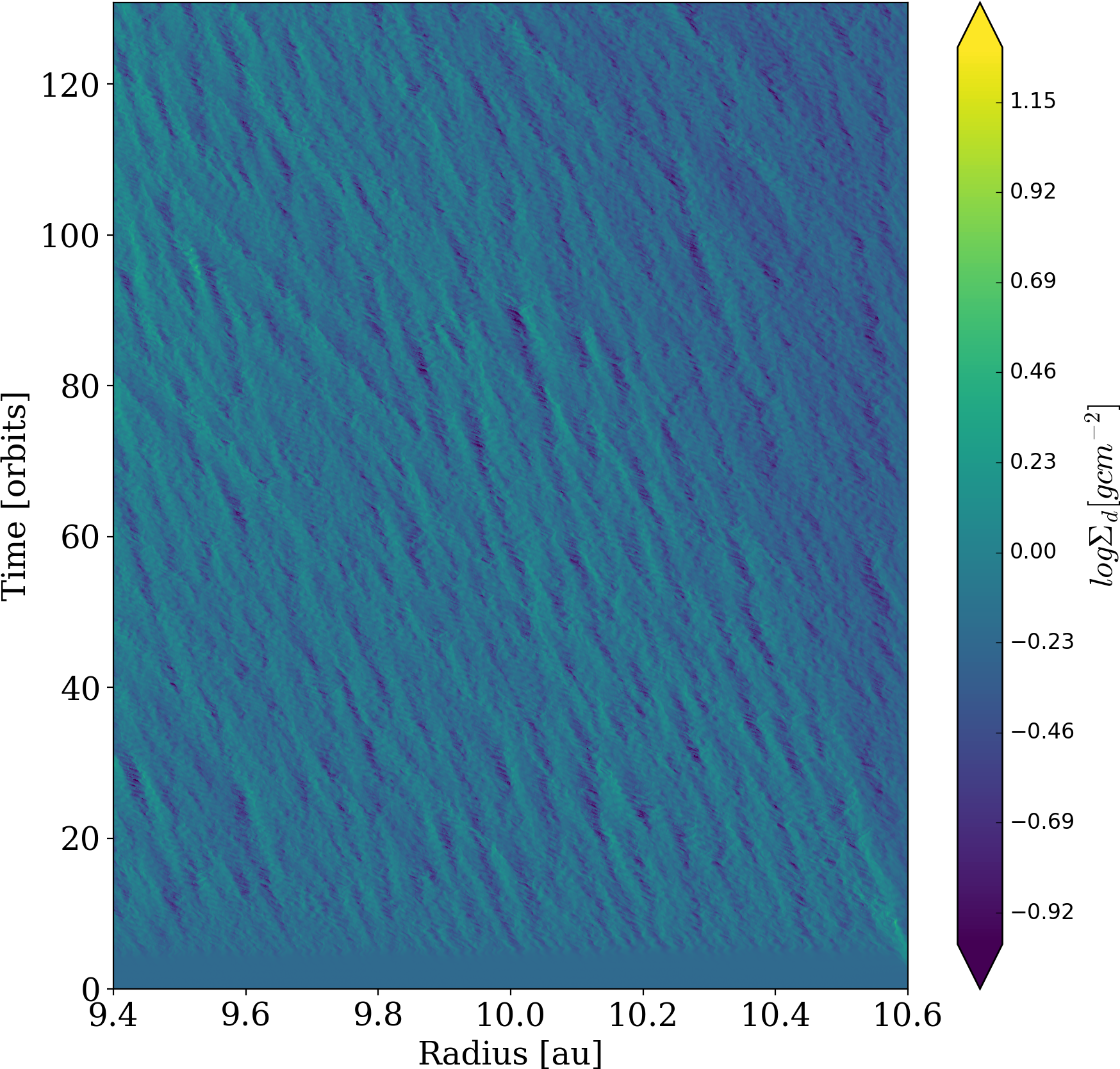}}
    \caption{Temporal evolution of the dust surface density over time for model \texttt{ST1} (top) and model \texttt{ST2} (bottom).}
    \label{fig:dsurfst1}
\end{figure} 

After a few orbits, the dust settles to the midplane and radially drifts further inwards. 
At the same time, the SI is triggered leading to dust concentration and clumping. Dust grains are resupplied in the outer buffer zones, enabling a constant pebble flux. In the following we investigate for the dust concentrations and the dust scale height in our models.

\subsection{Dust concentration and streaming instability}
 
In the following we analyse the maximum dust density at the center of the domain at $10$ au in a small radial patch of $0.1$ au. 
Results, plotted in Fig.~\ref{fig:dgrowth}, show that after roughly $10$ orbits the dust concentration reaches up to hundred times the initial value. 
Grains with $St=0.1$ (top panel in Fig.~\ref{fig:dgrowth}) show a strong concentration, attaining a (temporally-averaged) maximal dust to gas mass ratios of $\epsilon_{\rm max}=24 \pm 14$. 
The average concentration level at the midplane saturates at the time-averaged value of $\epsilon=4$ with a large scatter. 
Grains with $St=0.01$ (bottom panel in Fig.~\ref{fig:dgrowth}) show  slightly lower concentrations with (temporally-averaged) maximum concentrations of around $10 \pm 3$ of the dust to gas mass ratio. 
The spatial averaged midplane value of the dust to gas mass ratio is $2.8$. 


\begin{figure*}[t]
    \resizebox{\hsize}{!}{\includegraphics{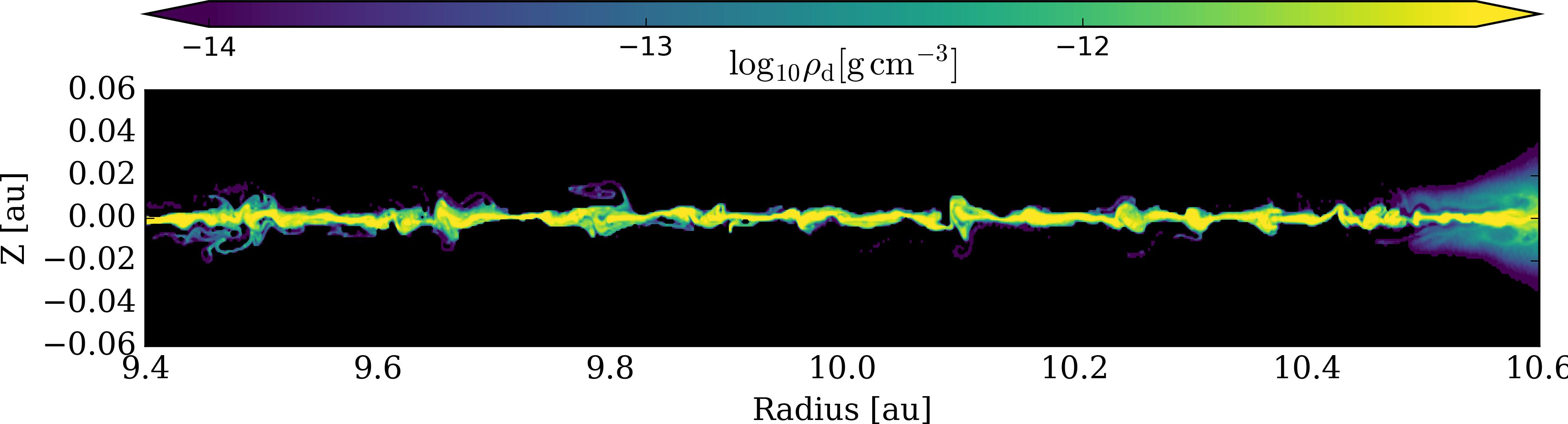}}
\resizebox{\hsize}{!}{\includegraphics{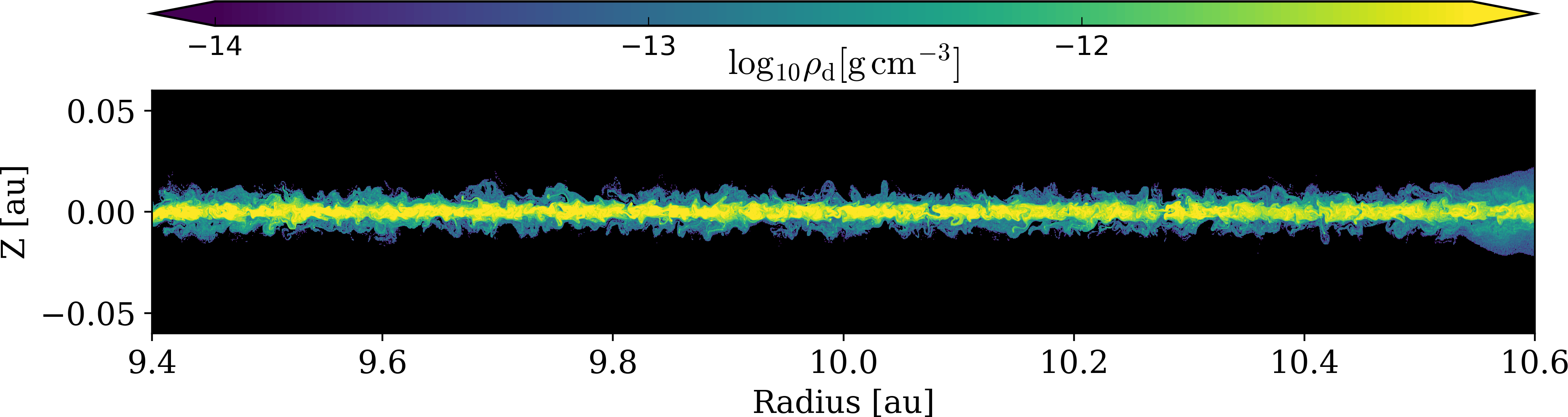}}
    \caption{Distribution of the dust density after 100 orbits for model \texttt{ST1} (top) and model \texttt{ST2} (bottom).}
    \label{fig:dust}
\end{figure*}

\subsection{Surface density evolution}
Fig.~\ref{fig:dsurfst1} shows the evolution of the vertically-integrated dust surface density for both models. 
After $10$ orbits, the SI transforms the smooth surface density into dust fragments of low and high dense filaments.
{ Fig.~\ref{fig:dsurfst1}, top, shows narrow, close to horizontal stripes which indicate the fast inward radial drift for model \texttt{ST1}.} At $\sim 80$ and $\sim 100$ orbits, two large dust accumulation become visible in the surface density which leads to a reduction of the radial drift. 
This is to be expected since large dust clumps can shield each other from the gas headwind.
In these large dust clumps, the maximum dust concentration can lead to dust to gas mass ratios of $\sim 100$, see Fig.~\ref{fig:dgrowth}, top, close to $90$ orbits. 

In the bottom panel of Fig.~\ref{fig:dsurfst1} we display the temporal evolution of the dust surface density for model \texttt{ST2}. 
Here the SI also leads to overdense structures in the surface density. 
Due to the slower inward radial drift, the dust clumps show a broader structure over time. 
The concentrations in the surface density remain on a similar level as model \texttt{ST1} although we do not observe the large dust accumulations. 


Fig.~\ref{fig:dust} shows snapshots of the dust density after $\sim 100$ orbits for both models in the meridional plane. 
The dust layer is very thin, with a vertical extent of only $0.01$ au and attaining dust densities of around $10^{-11}$~g cm$^{-3}$. 
Fig.~\ref{fig:dust} (top panel) shows dust clumps for model \texttt{ST1} on top of the narrow dust layer that remains concentrated in a region corresponding to roughly 1\% of the gas scale height. 
The turbulent structures have sizes of around $1/10$ of the gas scale height in radius, while there is a sharp density contrast along the vertical direction. 
Likewise we show (in the bottom panel of the same figure) the snapshot for model \texttt{ST2}. 
Here, the turbulent structures look much finer compared to model \texttt{ST1}, also with a smoother density contrast in the vertical direction. 
    

\subsection{Vertical dust scale height and effective $\alpha$}
\label{sec:dsh}

\begin{figure}
    \resizebox{\hsize}{!}{\includegraphics{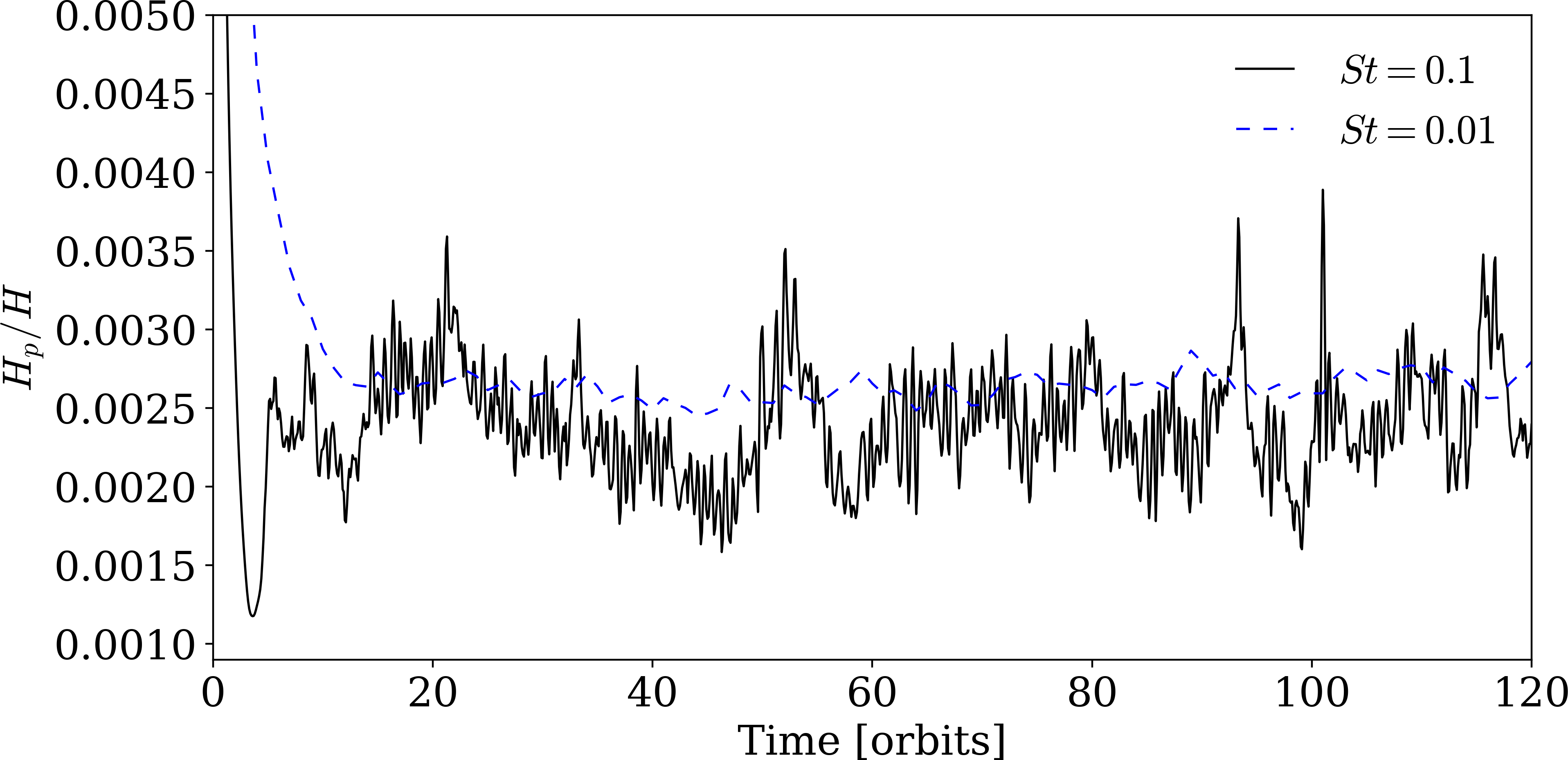}}
    \caption{Dust scale height evolution over time, shown for model \texttt{ST1} and model \texttt{ST2}.}
    \label{fig:Hp}
\end{figure}

To calculate the dust scale height we follow first the approach by \citet{yan17} and determine the standard deviation of the vertical position of the grains
\begin{equation}
  \frac{H_p}{H} = \frac{\sqrt{  \overline{z_p^2}-\overline{z_p}^2} }{H}\,,
\end{equation}
with $z=R \cos\theta$. 
Fig.~\ref{fig:Hp} shows the evolution of the dust scale height over time which remains between 0.2 to 0.3 \% of the gas scale height for both models.

\begin{figure}
    \resizebox{\hsize}{!}{\includegraphics{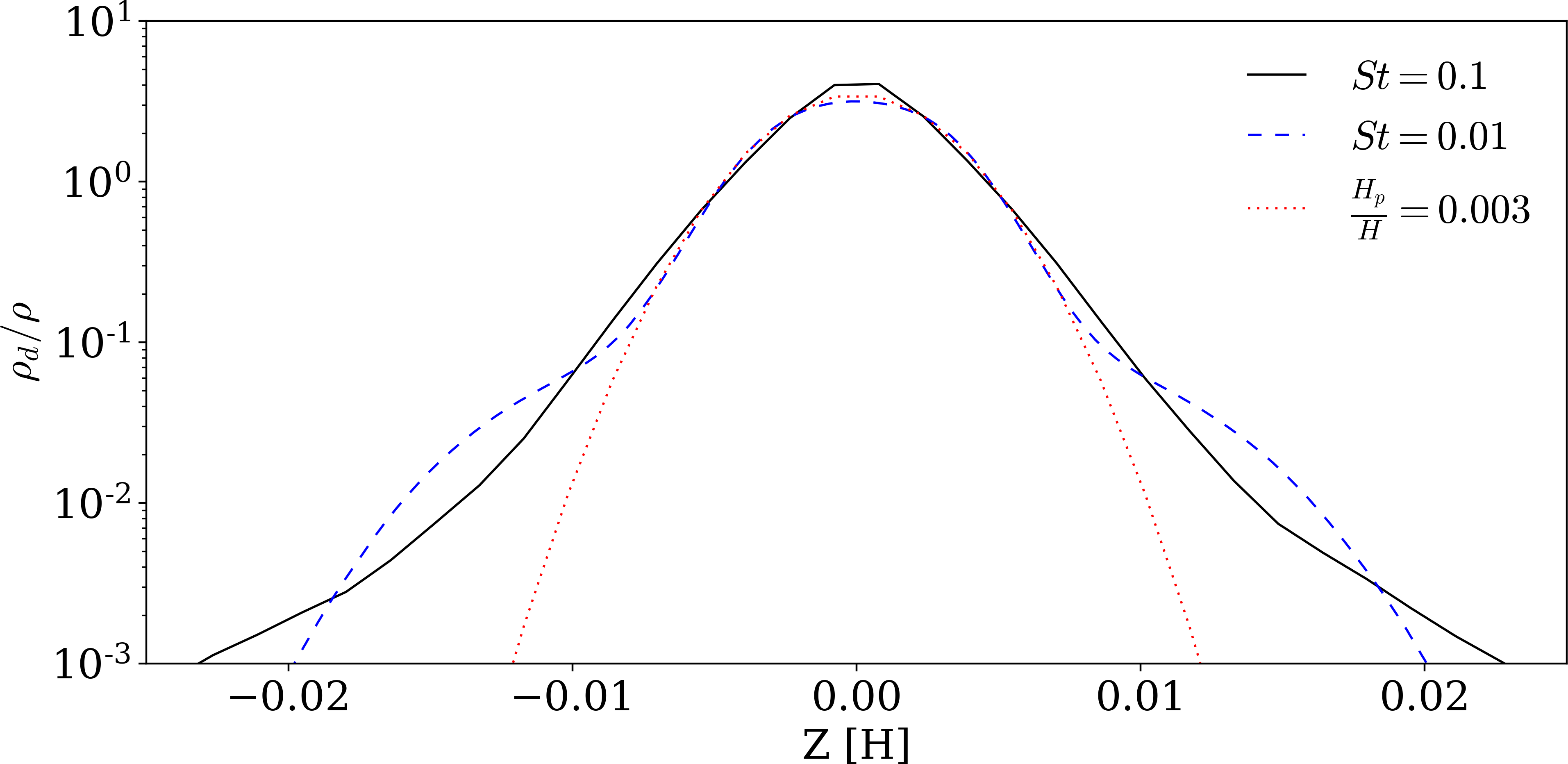}}
    \caption{Vertical profile of the dust to gas mass ratio, time and space averaged at 10 au, for model \texttt{ST1} and model \texttt{ST2}. A fitting profile is shown with the red dotted line.}
    \label{fig:Hp_vert}
\end{figure}

To verify the scale height determination technique we follow another approach outlined in \citet{flo20} and plot the averaged vertical dust density profile  
(see Fig.~\ref{fig:Hp_vert}).
The plot provides the time averaged profile 
from 20 orbits until the end of the simulation and spatially averaged at $10\rm \, au \pm 0.15$au.
The dust density is normalized by the gas density (which remains effectively flat in the vertical direction with only 5 per mill deviation). 
Both profiles fit best with a value $H_p/H$ of $\sim 0.003$, particularly inside the first 0.01 au from the midplane.
Above 0.01 au from the midplane the profile becomes more shallow, probably because of sudden bursts of dust clumps, as it is seen in Fig.~\ref{fig:dg}. 
By determining the dust scale height we can effectively determine the level of turbulence which would be equivalent to produce such a profile. 

Following the calculation from our previous work \citet{flo20} and \citet{dubrulle95} we can estimate the { turbulent diffusivity} $\alpha$ with

\begin{equation} \label{eq:hd}
  \alpha=\frac{ \mathit{St}\, \mathit{Sc}}{\frac{H^2}{H_p^2}-1}\,,
\end{equation}
assuming a Schmidt number $\mathit{Sc}$ of unity. 
Inserting the values for the Stokes numbers we derive an effective $\alpha$ of about $10^{-6}$ for model \texttt{ST1} and $10^{-7}$ for model \texttt{ST2}.

\section{Discussion} 
%

%

\subsection{Comparison to previous local box simulations}
\label{sec:comp}
In what follows, we compare our simulation results with previous stratified local box models of the SI. 
Two parameters are important to characterise the evolution of the SI, namely, the Stokes number and the value of $\sigma_{\rm d}=\Sigma_{\rm d}/(\rho_0 \eta r)$ which can be understood as an average dust-to-gas mass ratio at the midplane layer. Most of the dust mass is concentrated in a regions of 1\% of the gas scale height around the midplane, which motivates the need to quantify the SI using different parameters such as done by \citet{Sekiya2018} who adopted $\sigma_{\rm d}$. \citet{Sekiya2018} introduced the parameter $\sigma_{\rm d}$, based on 3D stratified local box simulations. 

\paragraph{Maximum dust-to-gas mass ratio}
Our models have $\rm St=0.1$ (\texttt{ST1}) and $\rm St=0.01$ (\texttt{ST2}) with $\sigma_{\rm d}=0.36$. 
\citet{Sekiya2018} presented a variety of simulation cases, the closest of which  are their models (A) having $(St=0.1,\, \sigma_{\rm d}=0.5)$, and model (G) with $\{St=0.01,\, \sigma_{\rm d}=0.5\}$.
While our model \texttt{ST1} favourably compares (in terms of the maximum dust concentration) to their model (A) ($\epsilon_{\max}\sim 24$ vs. $\epsilon_{\max} \sim10$), we found higher values for $\epsilon_{\max}$ for our model \texttt{ST2} compared to their model (G) ($\epsilon_{\max}=10$ vs. $\epsilon_{\max}=2.4$). 

\paragraph{Dust Scale Height}
In section~\ref{sec:dsh} we have shown that the dust scale height in steady state reaches values of $H_p/H=0.003$. { Because of the different convention of in local box simulations our H is a factor of $\sqrt{2}$ larger and $H = \sqrt{2} H^l$. To be able to compare with the previous local box simulations we include this factor which give $H_p/H^l=0.0042$.}

\citet{yan17} found a dust vertical scale height for $St=0.01$ grains and  $\sigma_{\rm d}=1$ of about $H_{p}=0.014$, { roughly 3 times} higher than in our models. 
\citet{carrera:2015} presented a model with $\sigma_{\rm d}=0.5$ and his particle scale height was around $H_p=0.005$, { very similar than the value we found. }
\citet{Sekiya2018} found that the strength of the SI scales with $\sigma_{\rm d}$, a result also predicted from the analytical works of \citet{squ18,pan20} who demonstrated that the growth rate depends on the dust to gas mass ratio. 
As our models adopt a lower value of $\sigma_{\rm d}$, it might be that the strength of the SI is reduced. 

\paragraph{Dust Clumping}
\citet{yan17} found long lasting dust concentrations appearing after hundreds to thousands of orbits while \citet{Sekiya2018} noticed that such concentrations appear for values of $\sigma_{\rm d} \ge 1$. 
Model \texttt{ST1} showed two events of secondary dust concentration reaching values of 100 times the gas density which reduced the radial drift of the dust clump, albeit this concentration was not enough to reach the critical Roche density, see appendix D. 
On the other hand, model \texttt{ST2} showed no major dust concentration. 
We also note that such dust accumulations have been observed on timescales of several hundreds of orbits \citep{yan17}. 
The radial drift in combination with our limited radial domain extent does not allow us to trace individual grains for such a long time (we cannot follow individual grains for more than around 10 orbits in model \texttt{ST1}). 

We thus conclude that our simulation results show similar values of dust clumping as observed in local box simulations. 
We find a smaller value of the dust scale height and no secondary long lasting dust accumulation, both possibly because of our choice of $\sigma_{\rm d}$.

{ Finally we performed a local box simulation with the same setup as presented in \citet{yan17} in Appendix E, to compare our new numerical method to previous models. We report that the results of dust concentration and particle vertical mixing in our local box runs are very similar as found in previous works.}




\subsection{Resolving the streaming and other instabilities} 
\label{sec:res}
Both grid resolution and particle sampling are important to correctly represent and resolve the dust and gas interactions. 

Grid resolution is crucial in order to resolve the fastest growing modes of the SI.
\citet{squ18} pointed out that the wavenumber of the fastest growing mode for the SI follows roughly $k \eta r \sim 1/\mathit{St}$. 
In our model \texttt{ST1}, assuming $k \sim 1/\Delta x$ and $\eta \sim (H/R)^2 = 0.0049$ we obtain $H^2/ (R\Delta r) \sim 45$, therefore allowing us to resolve wavenumbers (ideally) up to $45$ using $640$ zones per scale height.
%
%
%
Here the fasting growing mode corresponding to $k \eta r \sim 10$ is well resolved. 
Model \texttt{ST2} can capture wavenumbers $k \eta r$ up to { 180 and so also resolves the fastest growing mode corresponding to $k \eta r \sim 100$}. 
\citet{yan17} showed SI operating with $\mathit{St}=0.01$ with a range of resolutions from $k \eta r =32$ up to 256 while the strongest concentrations appeared when using resolutions close to the fastest growing wavelength. 

Accurate particle sampling is fundamental to capture correctly the dust feedback and to resolve the dust distribution \citep{mig19}.
\citet{Sekiya2018} adopt a particles over grid cells ratio $N_{par}/N_{cell} \approx 1.39$ while \citet{yan17} used a ratio of unity. 
In our models we employ $N_{\rm par}/N_{\rm cell} \approx 79.3$ for model \texttt{ST1} and $N_{\rm par}/N_{\rm cell} \approx 9.9$ for model \texttt{ST2}, both much larger than the previous models. 
In the appendix of \citet{yan17} he compared results obtained using $N_{\rm par}/N_{\rm cell} = 10$ and $N_{\rm par}/N_{\rm cell} = 1$ without finding any significant difference.

From this perspective, our models provide the necessary grid resolution and particle sampling in order to capture the basic SI properties as well as to resolve for the dust feedback. 

Another type of instability which might have an important effect is the vertical shear SI \citep{ish09} which is driven by the vertical gradient of the velocity shear between the dust and the gas. 
A recent work by \citet{lin2020} emphasizes the role of this instability in determining the vertical scale height of the dust. 
However, we point out that scales of the order of $10^{-3}H$ have to be resolved to capture the vertical shear SI.
Future high resolution simulations reaching ten thousand of cells per H are needed to verify the importance of new types of instabilities, like the settling instability or the vertical shear streaming instability.




\subsection{Gas transport}

Without damping, the gas surface density is quickly reduced creating strong radial pressure gradients which affected the gas surface density structure. 
Owing to the limited domain extent, we could not investigate this interesting effect. 
Specially due to the small vertical domain this effect is enhanced as there is no resupply of gas material from the upper layers. 
For this models we applied the gas density relaxation to study in more detail the SI in quasi steady state. 
Future simulations should include a much larger vertical extent to examine the effect of the dust drag onto the gas, particularly at regions where dust is expected to accumulate, such as the water ice line, where the dust drag can become very important for the gas motion \citep{Gar20}.



\subsection{Regions of planetesimal formation}

Over the recent years, several works have shown that the generation of planetesimal via the SI in a smooth disk profile remains difficult. 
First, a large amount of solid material (Z > 0.02), larger than the typical ISM value, has to be provided in a single dust species of certain Stokes number \citep{yan17,Johansen:2014}, leaving a relative narrow range of parameter \citep{umurhan:2020,che20}, even more narrow when including the effect of turbulence \citep{jaupart:2020}.
However such favorable conditions - a low amount of turbulence and a narrow range in mass distribution of large grains - is not expected from dust evolution models \citep{brauer:2008,birnstiel:2011}. 
Another difficulty arises in multi-grain simulations including different grain sizes, which showed the reduction of the SI grow regime \citep{krapp:2019}. 

On the other hand, the locations of pressure maxima in the disk remain plausible regions for planetesimal formation, owing to the large amount of dust concentrations and the conditions for the SI to operate are favourable \citep{Auffinger:2018,abod:2019,carrera:2020}. 

\subsection{The importance of the pebble flux}
\label{sec:pebl}
With this work we intend to emphasize the importance of the radial flux of pebbles rather than adopting the (more common) total gas-to-dust mass ratio when modeling the evolution of the SI. { The pebble flux controls the transport of the solid material in the disk, it can show us where dust grains get concentrated and trapped and it is important for the accretion of solid material onto planets \citep{orm18} which requires the understanding of their vertical distribution \citep{lai20}. The radial flux of pebbles can be determined using dust growth and evolution models evolution \citep{birnstiel12,takeuchi02,dra2016,dra21}.}

\begin{figure}
    \resizebox{\hsize}{!}{\includegraphics{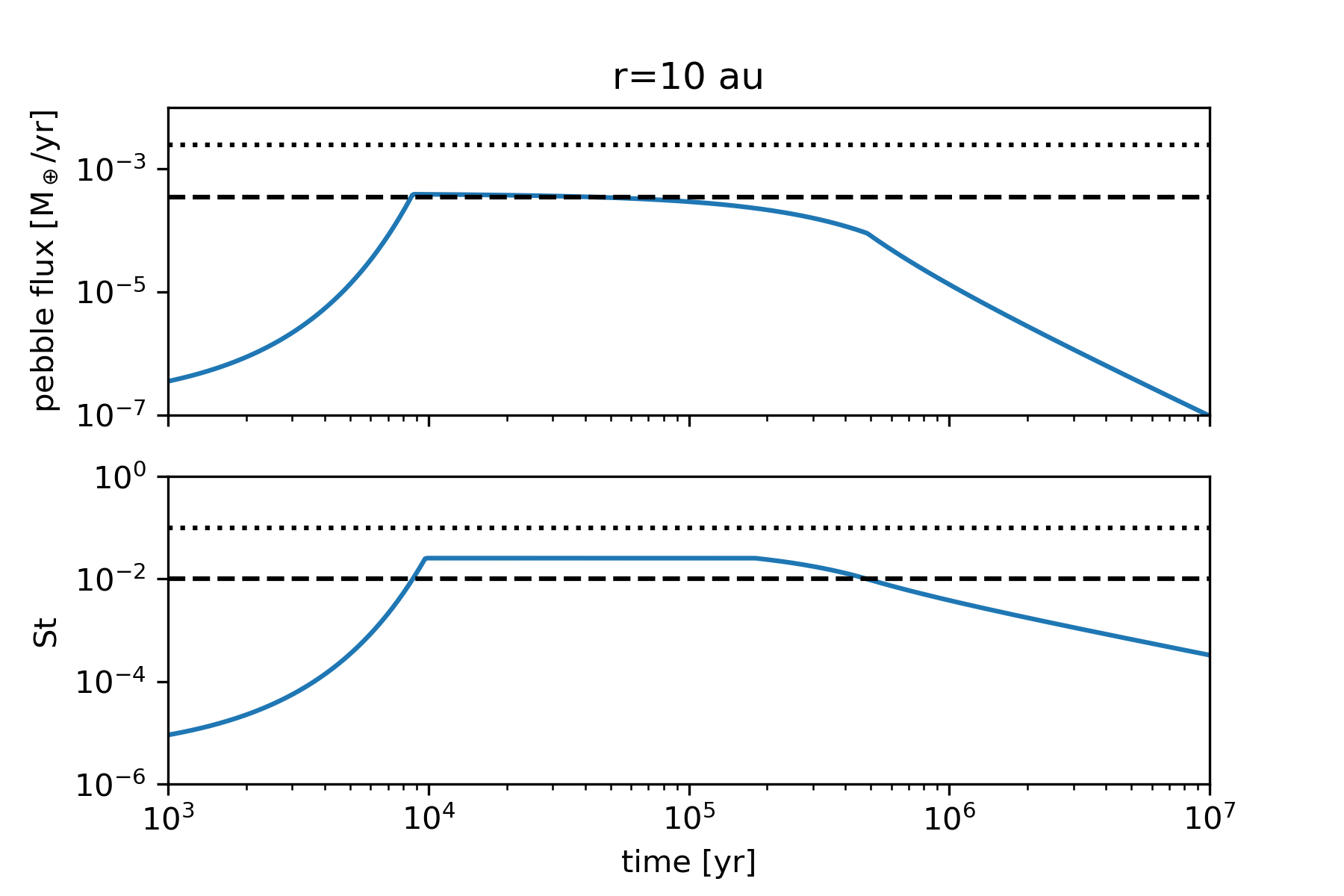}}
    \caption{Pebble flux over time, calculated with the pebble predictor tool using the same disk initial conditions. Overplotted are the values from our simulation results, \texttt{ST1} (dotted line) and model \texttt{ST2} (dashed line).}
    \label{fig:pebble}
\end{figure}

Using the pebble flux simulator\footnote{\href{https://zenodo.org/record/4383154\#.YBE-TZzPwWo}{Pebble predictor tool on Zenodo}} we determine the pebble flux over time using the same disk profile employed in our simulations. { The method and further references of the tool can be found in \citet{dra21}. The results are shown in Fig.~\ref{fig:pebble}.} 
The maximum pebble flux is reached at around $10^4$ years and it matches the value obtained in our model \texttt{ST2}, approximately $\dot{M}_{\rm d} \sim 3.5 \times 10^{-4} M_{\rm Earth}/year$ (see Fig.~\ref{fig:pebble} top). 
The pebble flux and grain sizes in model \texttt{ST1} lie above the predicted values from the dust evolution models.

We then conclude that our model \texttt{ST2} using $(St=0.01,\, \sigma_{\rm d}=0.36)$ presents realistic conditions of the dust amount when compared to models of dust evolution.
However such initial conditions are not favorable for secondary dust clumping events by the SI \citep{Sekiya2018} needed to account for planetesimal formation \citep{yan17}. 
This is an important aspect which should be investigated in more detail in forthcoming simulations of the SI.
%

\subsection{Total optical depth at mm wavelengths}
\begin{figure}
    \resizebox{\hsize}{!}{\includegraphics{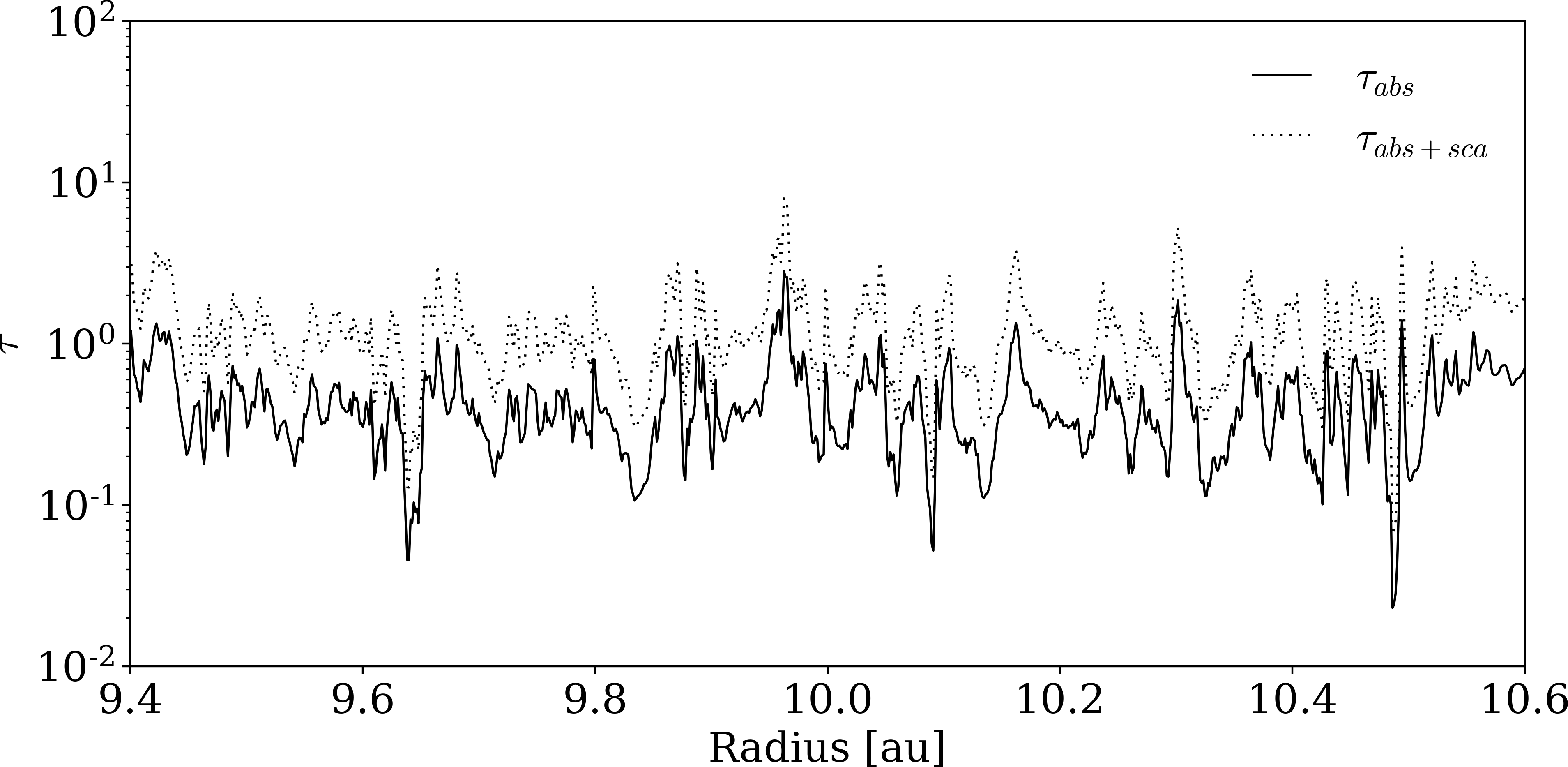}}
    \resizebox{\hsize}{!}{\includegraphics{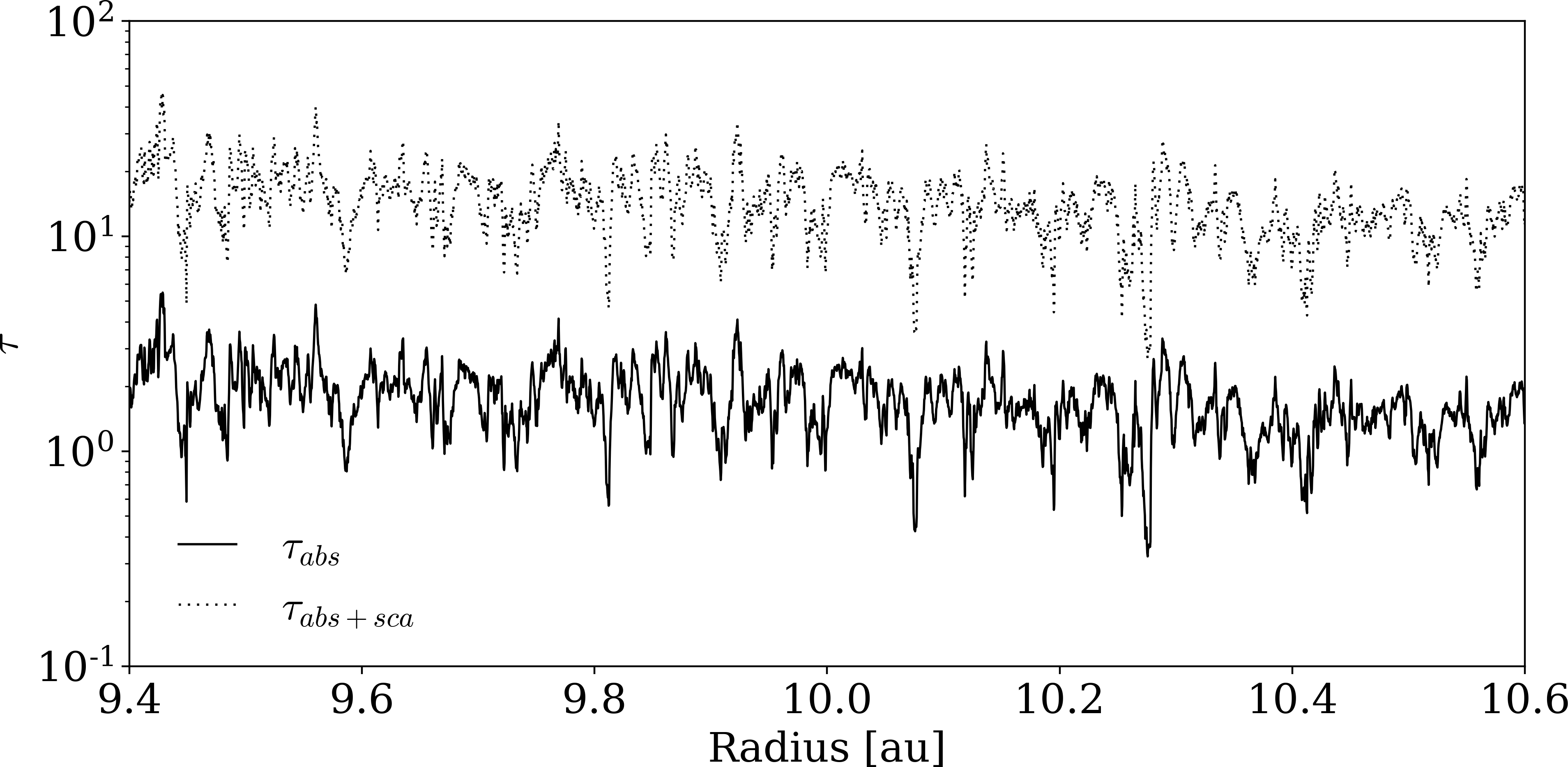}}
    \caption{Optical depth $\tau=\Sigma \kappa$ over radius shown for model \texttt{ST1} (top) and \texttt{ST2} (bottom). }
    \label{fig:tau}
\end{figure}

An important question remains whether the dust layer observed in protoplanetary disks are optical thin or thick at a given wavelength. 
For this, we calculate the opacity of the two grain sizes at the wavelength of $\lambda=1.3 \, {\rm mm}$ corresponding to the ALMA Band 6 observations. 
To calculate the opacity we use the optool\footnote{https://github.com/cdominik/optool/} { which is using the DIANA dust properties \citep{toon:1981,woitke:2016} and including the distribution of hollow spheres method \citep{min:2005} to calculate the dust opacity}. 
For the specific settings we use amorphous pyroxene (70\% Mg) with a mass fraction of 87 \% and 13 \% of amorphous carbon \citep{zubko96,preibisch:1993} and a water ice mantel with a mass fraction of 20 \% and a porosity of 20\%. 
We calculate the opacity for the two grain sizes using a narrow-size bin (0.8mm to 1mm) and (0.8cm to 1cm) for model \texttt{ST2} and model \texttt{ST1} respectively. 
The corresponding absorption and scatter opacity at $\lambda=1.3\,{\rm mm}$ are $\kappa_{abs}=3.009\, {\rm cm}^2/{\rm g}$, $\kappa_{scat}=21.957\, {\rm cm}^2/{\rm g}$ and for model \texttt{ST1} these are $\kappa_{abs}=0.635\, {\rm cm}^2/{\rm g}$ and $\kappa_{scat}=1.166\, {\rm cm}^2/{\rm g}$. In Fig.~\ref{fig:tau} we plot the radial profile of the total optical depth $\tau=\Sigma_{\rm d} \kappa$ calculated for both models using the vertical integrated dust density. 

The profiles show that for model \texttt{ST1}, the total optical depth remains around unity, while the optical depth from pure absorption opacity remains mostly optically thin.
For model \texttt{ST2}, the grain size is closer to the corresponding wavelength. 
Here the optical depth is larger and it remains mostly above unity. 
The scatter opacity is much larger for this grains which leads to a total optical depth of around $10$. More and more observations of protoplanetary disks at mm wavelength confirm the important effect of scattering \citep{sierra2020}. Also grain sizes of around mm size are consistent with the observations \citep{car19}. 

Overall, the variations in $\tau_{\rm abs}$ caused by the SI fluctuate between $0.4$ to $4$ (thus a factor of $\sim10$) in model \texttt{ST2} and between $0.02$ and $2$ (a factor $\sim 100$) for model \texttt{ST1}.
We point out again that these structures are on spatial scales of tens of $H$, which translates to scales of 0.1 au at the distance of 10 au from the star. 
Current radio interferometer capabilities of ALMA reach a spatial resolution of $5$ au for the dust emission at mm wavelength in the most nearby star disk systems.
 
\section{Conclusions}
%
%
In this work, we have presented a new generation of models to investigate the dust and gas drag instabilities in global stratified simulations of protoplanetary disks. 
High resolution, 2D global hydrodynamical simulations have been performed,  including the dust back-reaction on the gas modeling the conditions of a protoplanetary disk around a one solar mass star. 
Our numerical method is based on the hybrid fluid-particle framework recently developed by \cite{mig19}, where the dust component is modeled by Lagrangian particles.
We adopt 2D spherical geometry covering the meridional domain $(r:\theta)$ with a grid resolution up to 1280 cells per gas scale height to resolve for the streaming instability. 
The dust grains are modeled with a constant Stokes number of $St=0.1$ and $St=0.01$ which corresponds to grain sizes of 880 micron and $8.8$ mm,  respectively, at $10$ au.
The dust grains radially drift through the domain and they undergo streaming instability, leading to the formation of large dust concentrations. 
By resupplying dust grains at the outer radial domain, we reach a quasi-steady state of pebbles flux and operating streaming instability. Our main results may be summarized as follows:

\begin{itemize}
    \item The streaming instability leads to dust clumping, with maximum values between $10$ to $100$ in terms of the dust to gas mass ratio. 
    The average dust to gas mass ratio at the midplane remains between $2$ and $4$.   
   \item For $St=0.1$ we observe the appearance of large dust clumping reaching dust to gas mass ratios above $100$ which can effectively reduce the radial drift as grains shield itself from the gas drag.
   \item We found that the dust layer remains concentrated within a region of $\pm$ 0.01 au around the midplane. 
   Our models show an effective dust scale height of about $H_p/H=0.003$ independent of the Stokes number.
   \item We reach a nominal flux of pebbles of $\dot{M}_{\rm d} \sim 3.5 \times 10^{-4}\, M_{\rm Earth}/{\rm year}$ ($\sim 2.5 \times 10^{-3}\, M_{\rm Earth}/{\rm year}$) for grains with $St=0.01$ ($St=0.1$). The grain size and pebble flux for model $St=0.01$ compares best with dust evolution models of the first million years of disk evolution.
\end{itemize}


We finally wish to emphasize the important role of the pebble flux when determining the amount of dust in simulations of the streaming instability.
The maximum pebble flux in the disk is reached during the first million years of disk evolution. 
T-Tauri star disk models with a pebbles flux of around $3.5 \times 10^{-4}$ Earth masses per year and Stokes numbers $0.01\lesssim St \lesssim 0.1$ are closed to what is expected from dust evolution models. 
For this range of parameters, $\sigma_{\rm d}$ remains below unity making secondary dust clumping for planetesimal formation difficult \citep{Sekiya2018}.
This novel class of global dust and gas simulations constitute a promising tool for forthcoming studies targeting gas and dust evolution in protoplanetary disks, especially for situations where the density and pressure are strongly changing (such as at pressure maxima).

{\it Acknowledgements:}\\ 
We thank Jonathan Squire for helpful comments on the manuscript. M.F. acknowledges funding from the European Research Council (ERC) under the European Union’s Horizon 2020 research and innovation program (grant agreement No. 757957). Figures where produced by python matplotlib library \citep{Hunter:2007}.

\bibliographystyle{aa}
\bibliography{references}

\appendix
\section*{A. Dust sampling}
%

\begin{figure}
    \resizebox{\hsize}{!}{\includegraphics{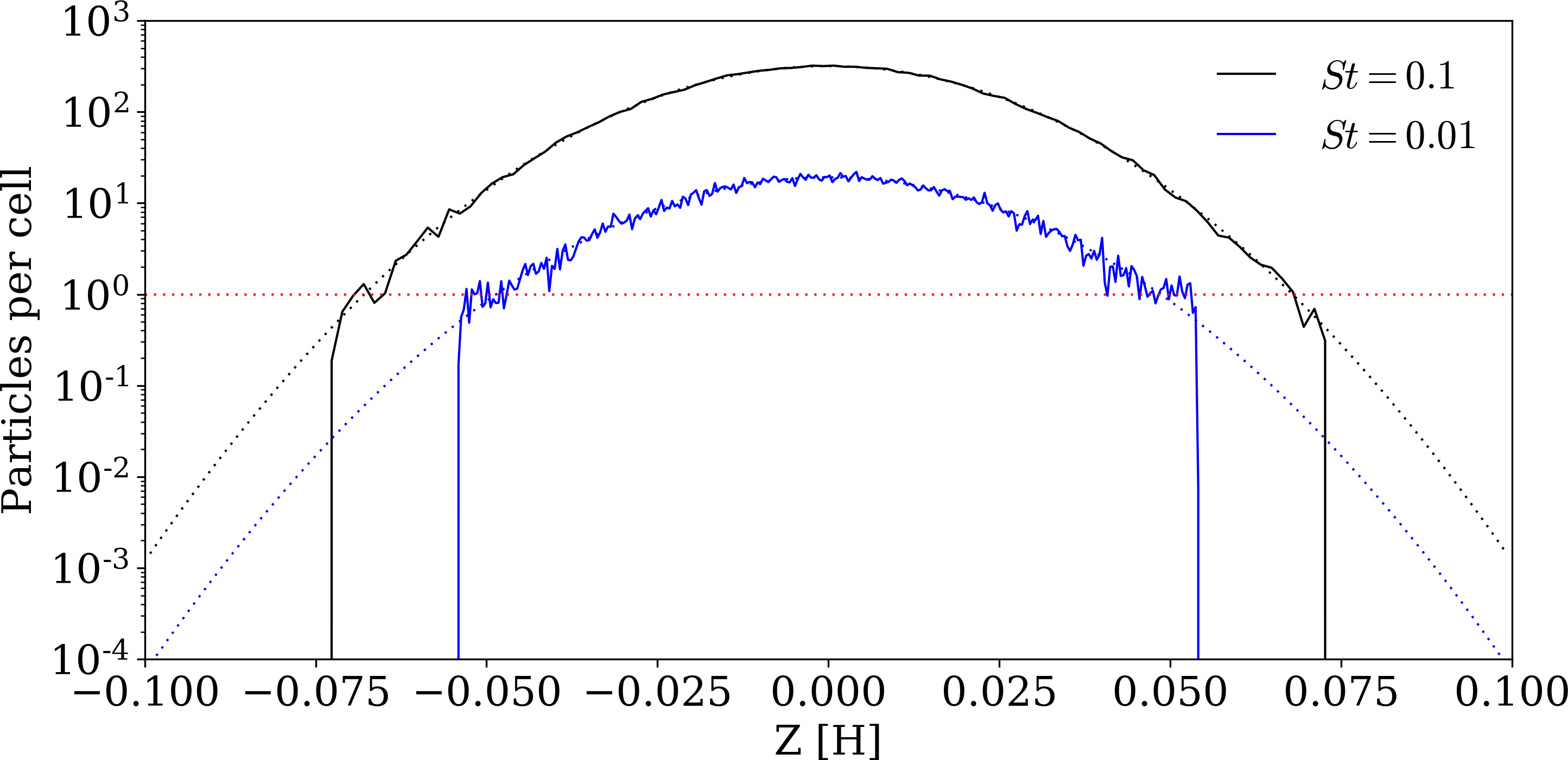}}
    \caption{Number of particles per cell along the vertical direction for both models. The red dotted line marks the position of 1 particle per cell.}
    \label{fig:sampling}
\end{figure}
In our models, the dust density is sampled by individual particles. In Fig.~\ref{fig:sampling} we show the number of particles per cell along the vertical direction for both models \texttt{ST1} and \texttt{ST2} at 10au. 
The black and blue dotted lines shows the initial dust density profile and the theoretical sampling. 
The dust density can only be sampled until a given height, where one has at least one particle per cell, as indicated with the red dotted line in Fig.~\ref{fig:sampling}.

%
\section*{B. Benchmark - Dust sampling}
%
\begin{figure}
    \resizebox{\hsize}{!}{\includegraphics{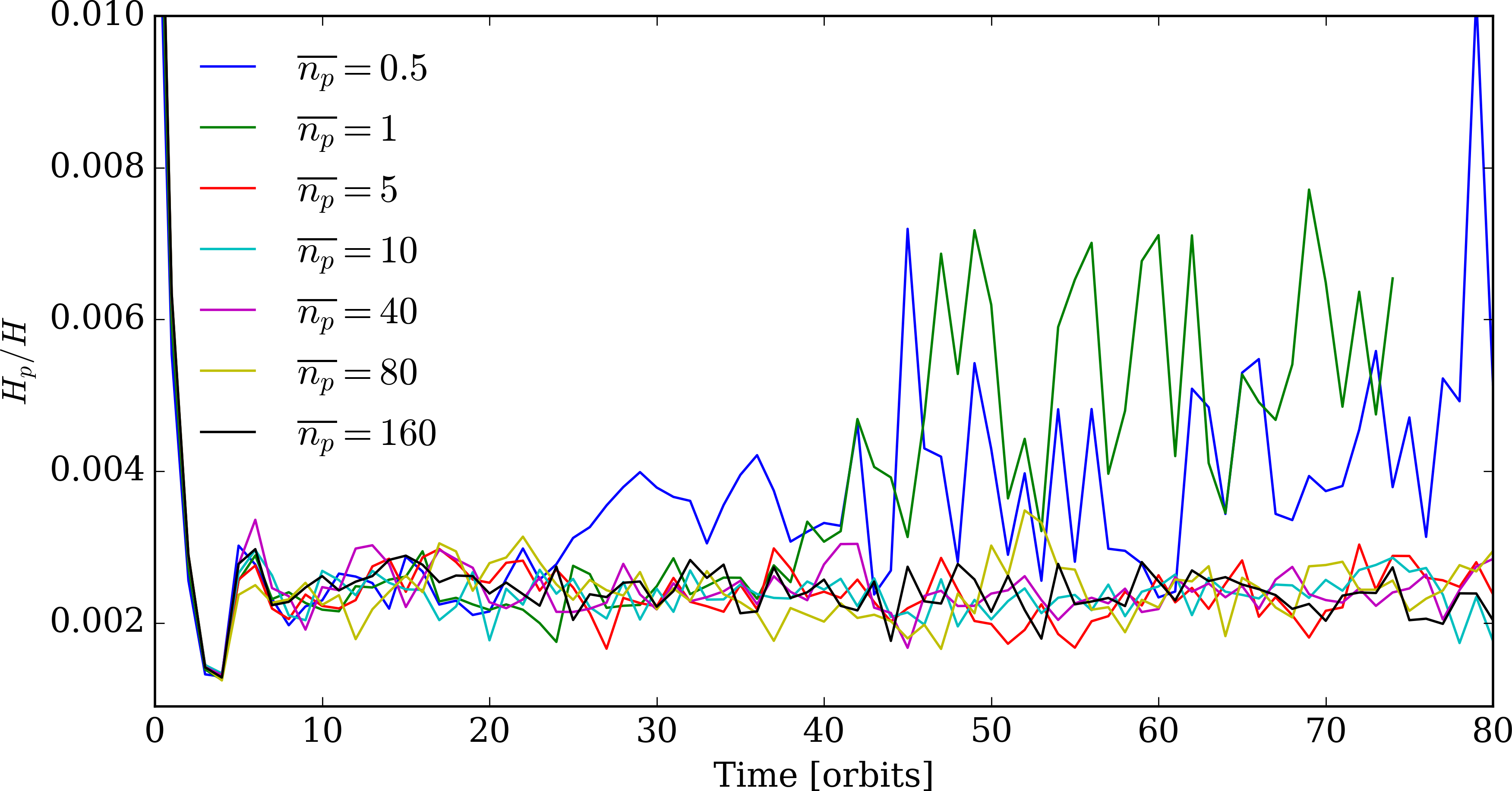}}\\
    \resizebox{\hsize}{!}{\includegraphics{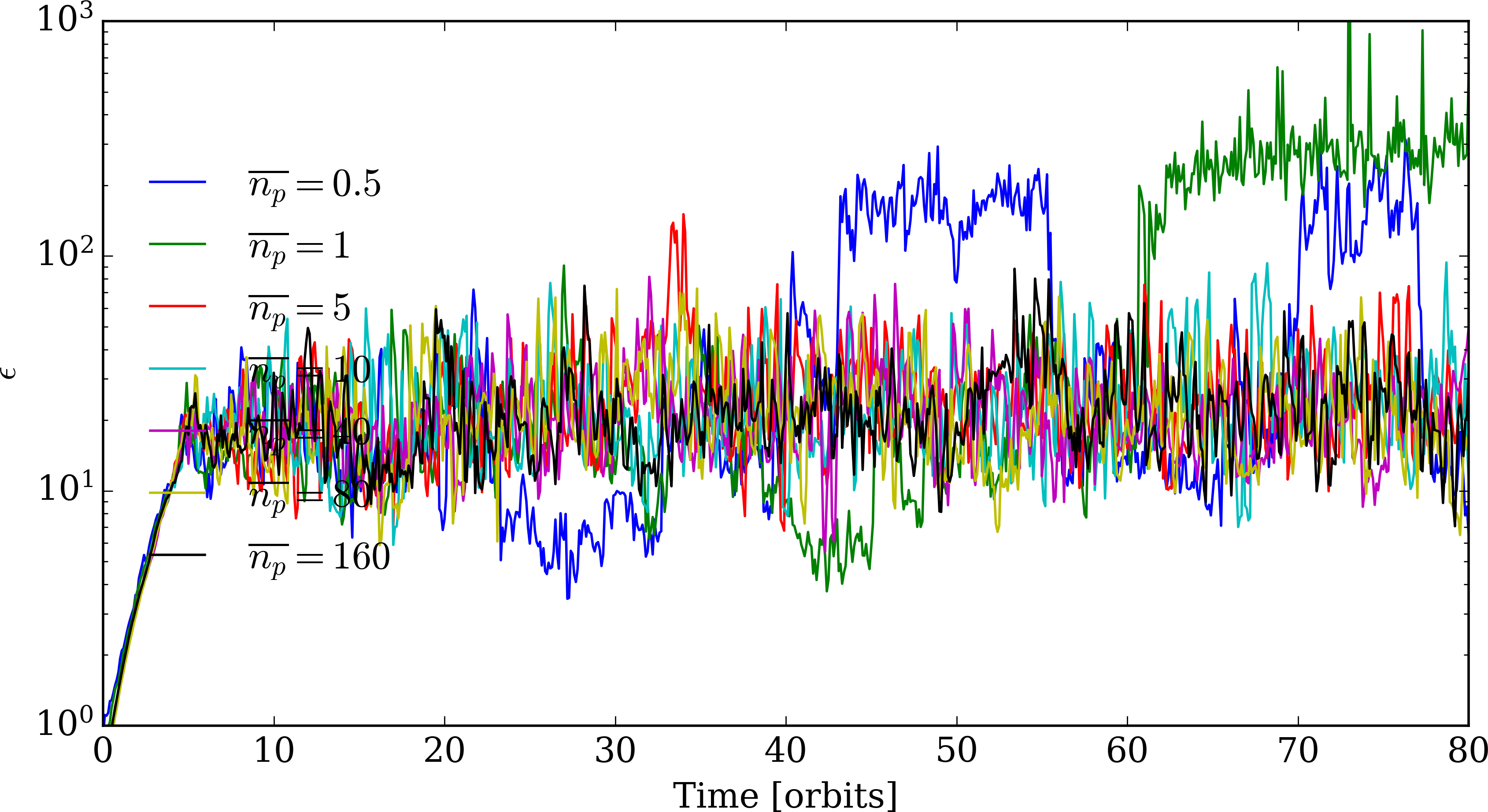}}
    \caption{Benchmark results of model \texttt{ST1} for different sampling rates, showing the dust scale height (top) and the maximum dust concentration at 10au.}
    \label{fig:bench}
\end{figure}
Here we investigate the effect of the particle sampling. 
For this we perform a series of runs, based on model \texttt{ST1}, using different numbers of particles, ranging from 0.5 up to 160 particles per cell. 
Fig.~\ref{fig:bench} shows the dust scale height and the maximum dust concentration $\epsilon_{\rm max}$ at 10 au over time. 
The results indicate that a sampling of 5 particles per cell or more is enough to show converging results. 
Below this value sudden dust concentrations occur which also trigger larger dust scale heights, possibly due to Kelvin-Helmholtz Type instabilities.

\section*{C. Gas damping}
%

\begin{figure}
    \resizebox{\hsize}{!}{\includegraphics{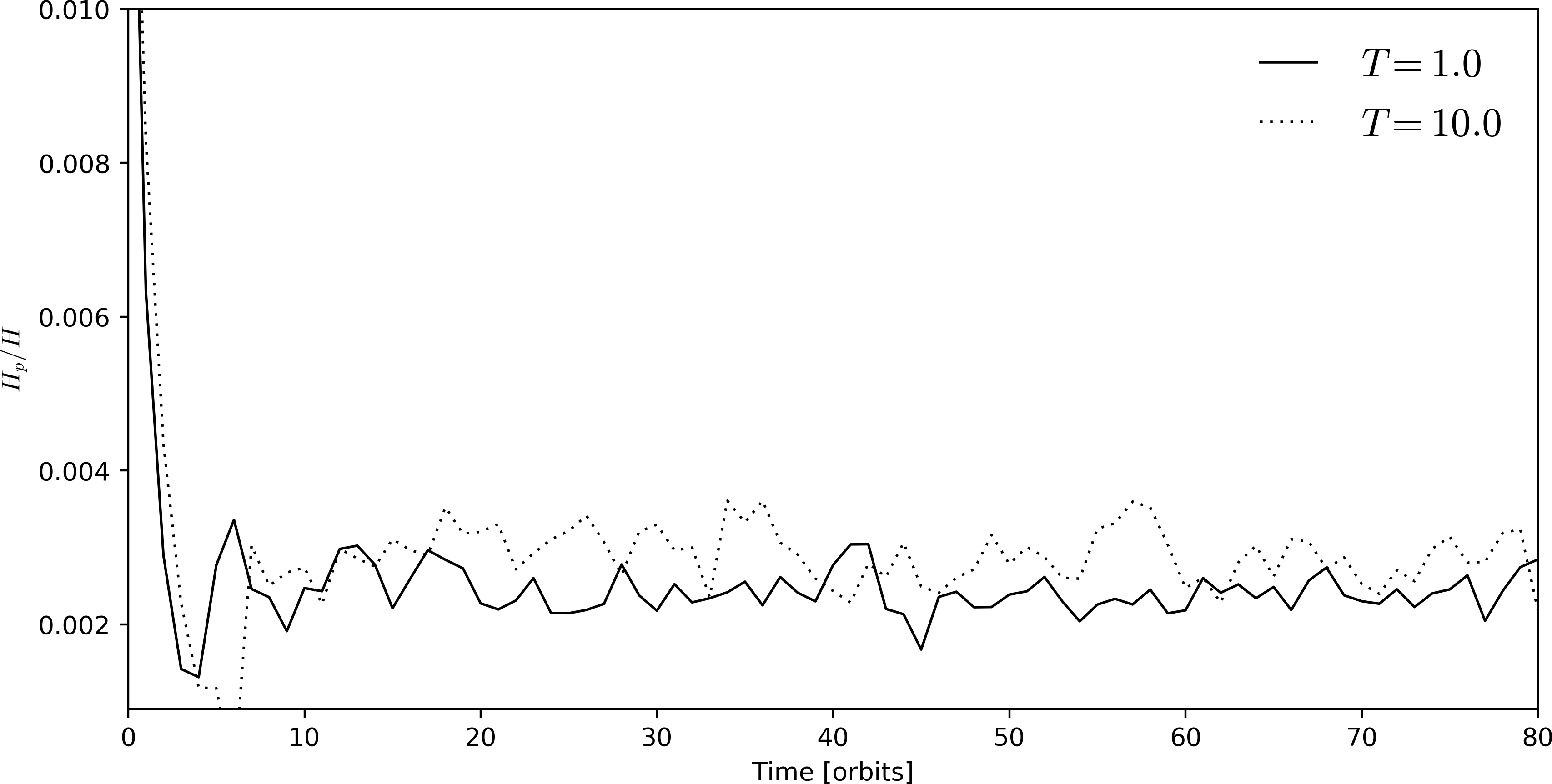}}\\
    \resizebox{\hsize}{!}{\includegraphics{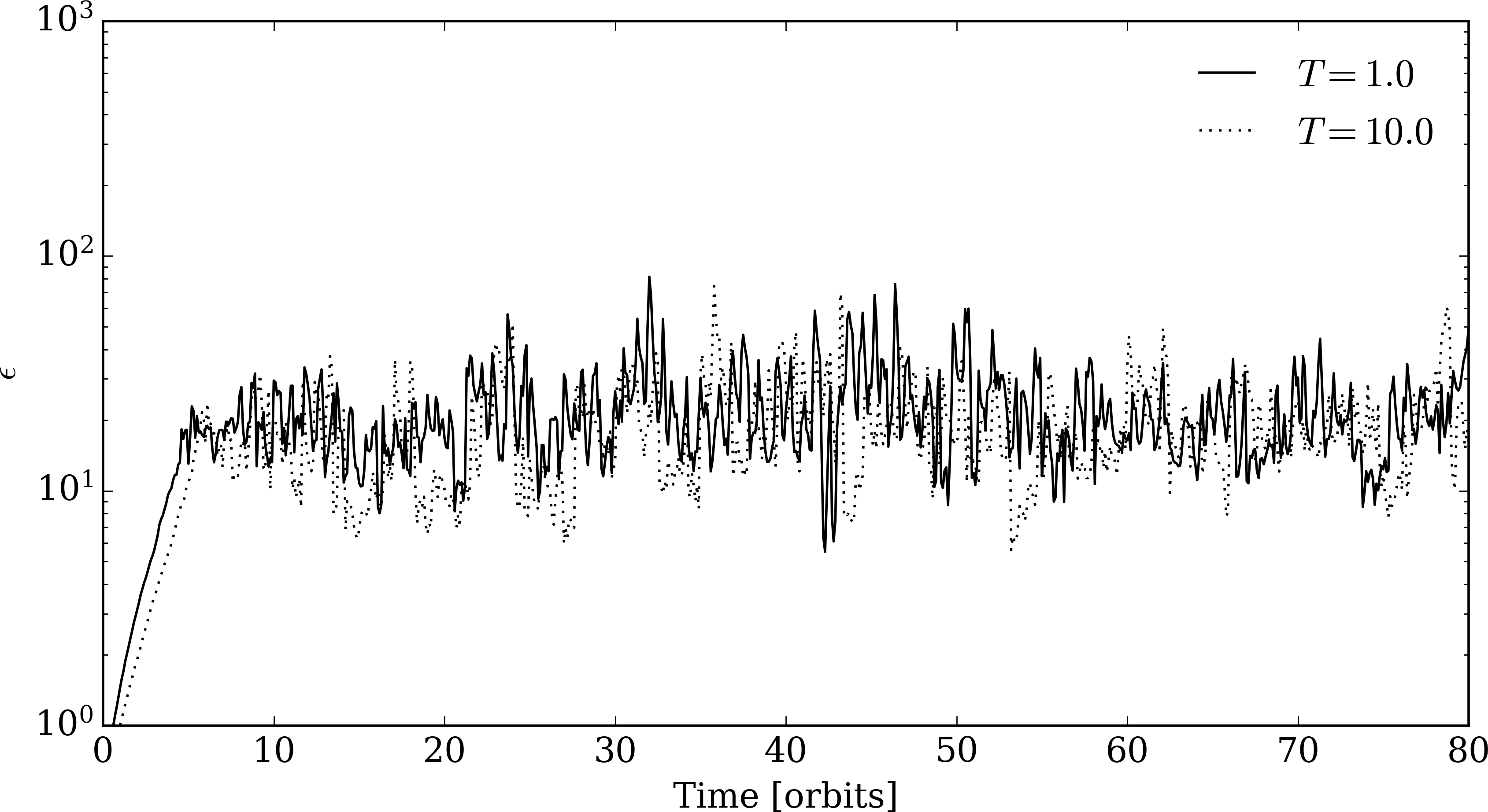}}\\
    \resizebox{\hsize}{!}{\includegraphics{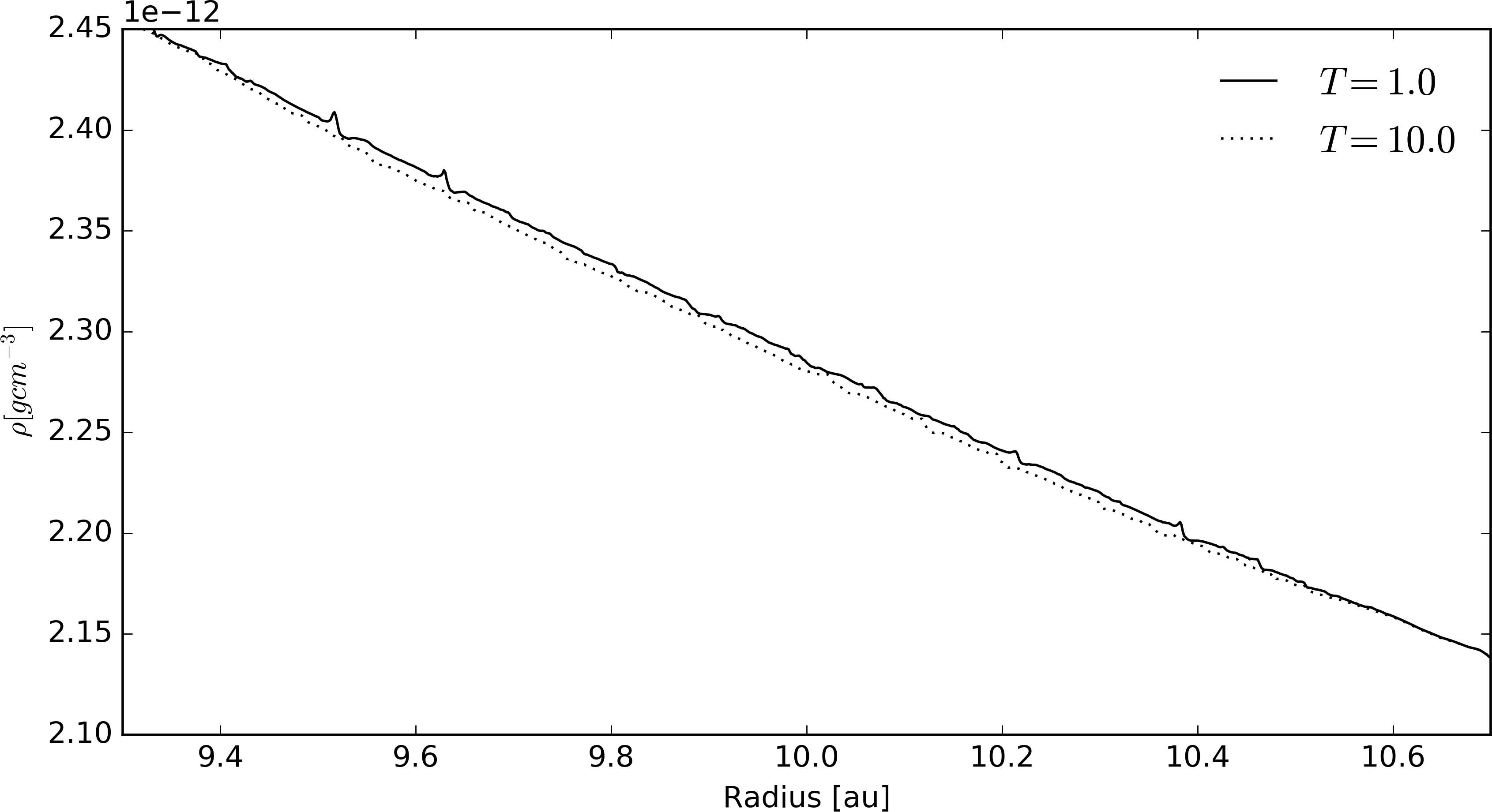}}
    \caption{Comparison results using two runs with different gas relaxation parameter for model \texttt{ST1}. Shown is the dust scale height (top), the maximum dust concentration (middle) and the midplane radial profile of the gas density.}
    \label{fig:bench_comp_damp}
\end{figure}
As we relax the gas density in our domain we have to verify that the gas damping does not strongly affect the non-linear evolution of the streaming instability. 
To this end, we perform a test run setting the damping factor $T=10.0$ in Eq.~\ref{eq:damp} for the gas damping, leading to a reduced gas damping rate. 
The results are summarized in Fig.~\ref{fig:bench_comp_damp}. The lower gas damping does not strongly effect the main results.

\section*{D. Roche Density}
%
\begin{figure}
    \resizebox{\hsize}{!}{\includegraphics{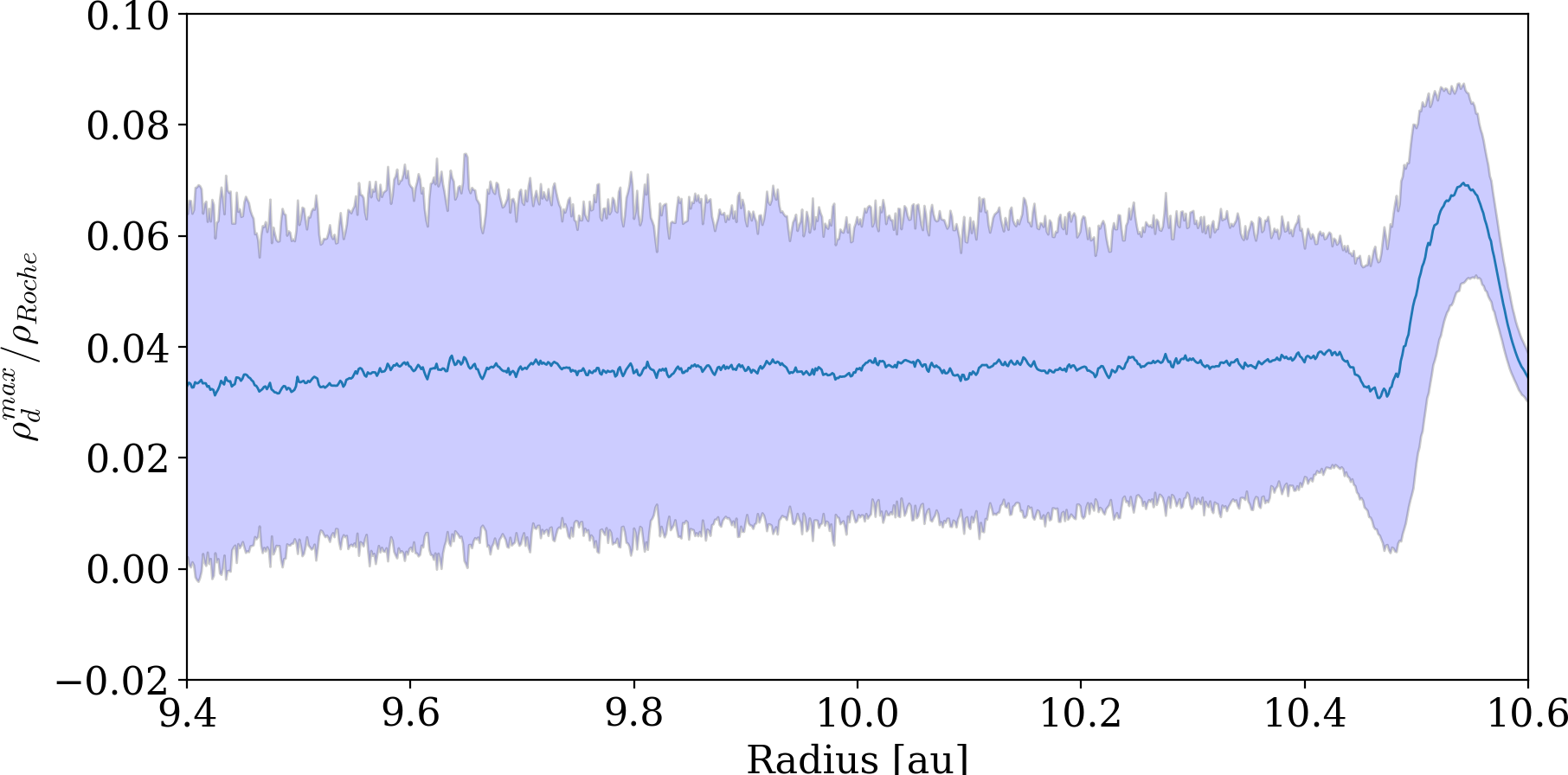}}
    \caption{Maximum dust density over radius, time averaged for model \texttt{ST1} and normalized over the Roche density. Solid line and filled area present the mean and standard deviation.}
    \label{fig:roche}
\end{figure} 

A common approach to investigate whether the SI could produce directly planetesimals through dust clumping which then collapse due to the self-gravity, is by determining the Roche density:
\begin{equation}
  \rho_{\rm Roche} = \frac{9}{4 \pi} \frac{M_*}{R^3}.
\end{equation}
In Fig.~\ref{fig:roche} we determine the maximum density (normalized to the Roche density) by computing the value at each radial position and then taking the average value over time. 
Fig.~\ref{fig:roche} shows that on average the maximum density concentration at the midplane reaches around $\sim 3\%$ of the Roche density. 
The profile remains very flat with a small rise close to $10.6$ au due to the injection of grains in the outer buffer zone. 
We note that for our models we never reached the Roche density in dust in a single cell.

\section*{E. ShearingBox}
%

\begin{figure*}
  \label{fig:sb_maps}
  \centering
  \includegraphics[width=0.44\textwidth]{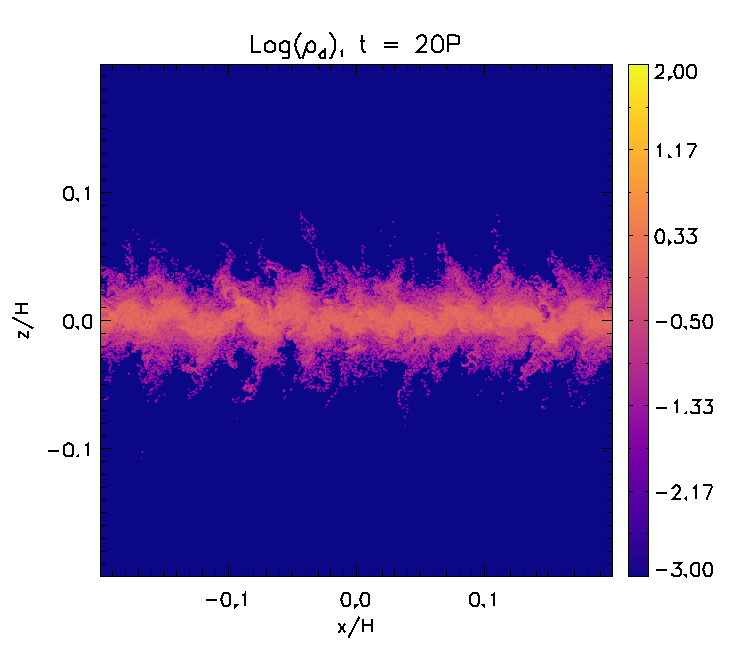}%
  \includegraphics[width=0.44\textwidth]{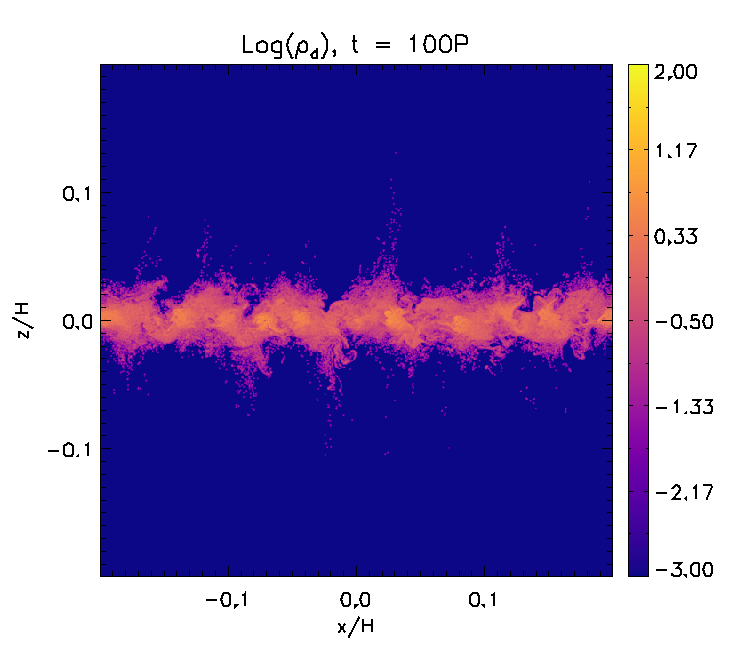}
  \includegraphics[width=0.44\textwidth]{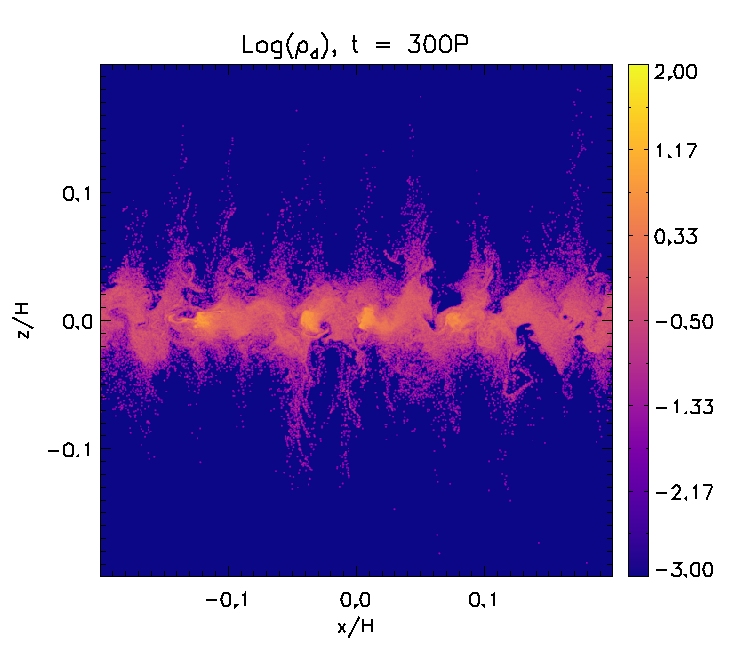}%
  \includegraphics[width=0.44\textwidth]{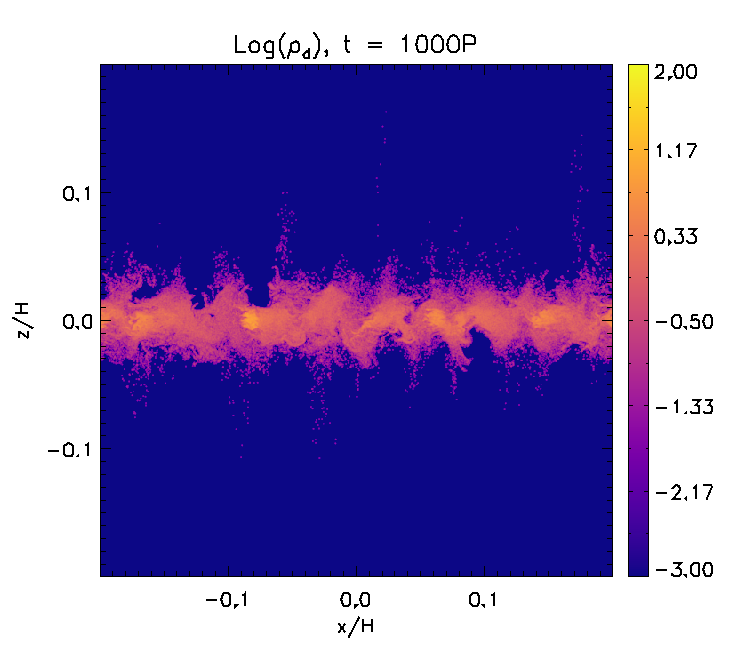}
    \caption{Dust density coloured maps at $t/P=20, 100$ (top panels)
             and $t/P = 300, 1000$ (bottom panels) for the shearingbox model}
\end{figure*} 

\begin{figure}
    \label{fig:sb_Hp}
    \resizebox{\hsize}{!}{\includegraphics{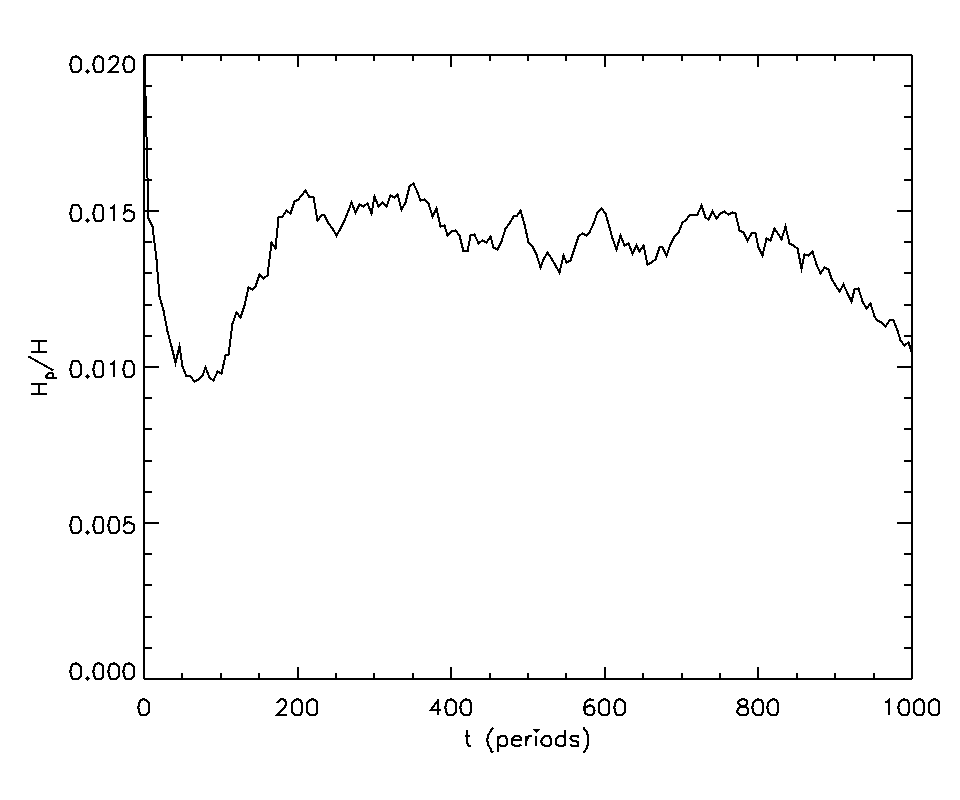}}
    \resizebox{\hsize}{!}{\includegraphics{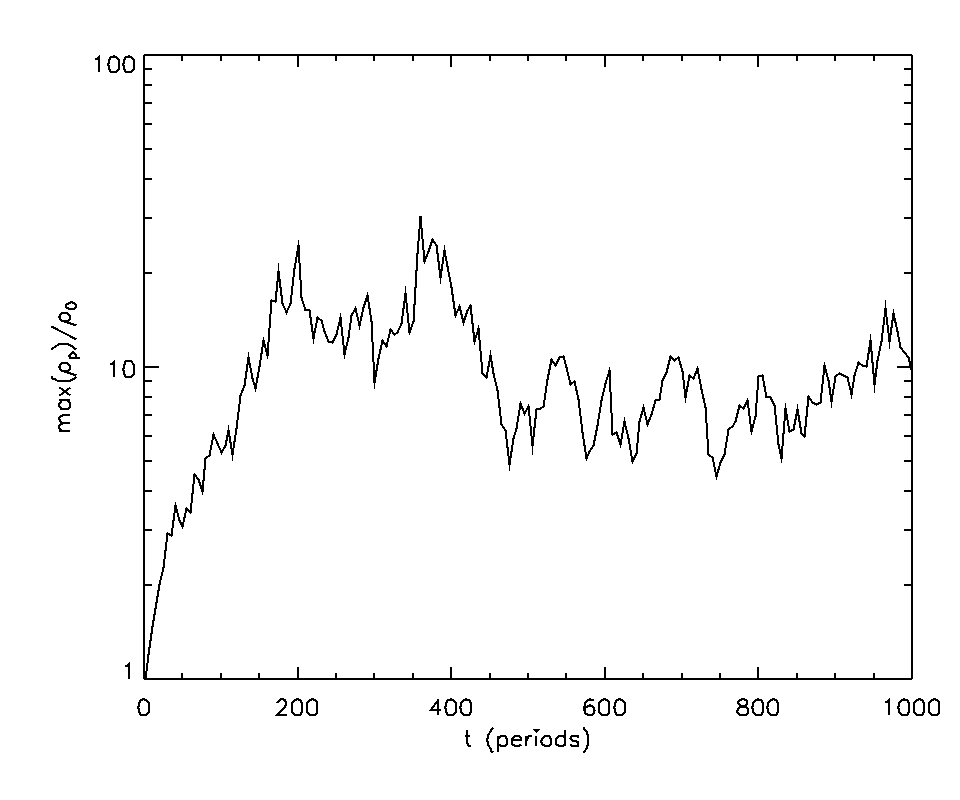}}
    \caption{Particle scale height as a function of time.}
\end{figure} 

{ 
In order to better compare our results to previous models of the streaming instability using local box simulations we performed a model which is similar to \cite{yan17}. In this model, we solve the axisymmetric shearingbox equations in the $(x,z)$ plane with $x,z\in[-0.2H,\,0.2H]$ where $H = c_s/\Omega$ is the vertical scale height.
The module has been thoroughly described in \cite{mig19}.

At $t=0$, we initialize the fluid state using the Nakagawa equilibrium \citep{nak86}:
\begin{equation}
  \vec{v}  = \displaystyle \frac{\eta v_K}{\Delta}\left[ 
       2\epsilon\tilde{\tau}_s,\,
     -\frac{\Delta + \epsilon\tilde{\tau}_s^2}{1+\epsilon},\,
      0\right]\,,
\end{equation}
where $\Delta = (1+\epsilon)^2 + \tau_s^2$, $\tilde{\tau}_s=\Omega\tau_s$
Here $\epsilon = 0.01=\rho_d/\rho_0$ and $\tau_s = 0.01$ are the dust to gas mass ratio and dust particles stopping time, respectively.
Gas density is initially set to unity ($\rho_0=1$) and an isothermal equation of state $p=\rho c_s^2$ is adopted, where $c_s$ is the sound speed.
Our units are chosen so that $\Omega = 1$ and $H=1$ (it naturally follows that $c_s = 1$).
The quantity $\eta v_K = 0.05c_s$ represents the external radial pressure gradient included on the gas.
As in \cite{yan17}, we neglect vertical gravity on the gas since no appreciable density stratification is present in the computational domain. 
We do, nevertheless, include linearized gravity ($g_z = -\Omega^2 z$) on the particles.

Dust grain velocities are also initialized with the Nakagawa equilibrium,
\begin{equation}
  \vec{v}_{p} = -\frac{\eta v_K}{\Delta}\left[2\tilde{\tau}_s,\, 
                  \frac{\Delta - \tilde{\tau}_s^2}{1+\epsilon},\, 0 \right],
\end{equation}
\citep[note that an incorrect factor $\epsilon$ appears in the expression for $\vec{v}_p$ in Eq. 55 of][]{mig19} while their position is assigned as
\begin{equation}
  \vec{x}_p = \left[x_b + (i+0.5)\Delta x,\, 0,\, z_p = r_g\right],
\end{equation}
where $x_b=-0.2H$ is the leftmost boundary, $i = 0,N_x-1$, $\Delta x$ is the mesh spacing along the $x$-direction and $r_g$ is a Gaussian random number with mean $\mu = 0$ and $\sigma = 0.02H$.
This mimics a spatial distribution of dust $\rho_d \sim \exp(-z^2/2\sigma^2)$ with reduced scale height in order to shorten  the sedimentation phase process as it was done in \citet{yan17}.

Particle mass is prescribed \citep[Eq. 1 of][]{yan17} according to:
\begin{equation}
  m_p = \sqrt{2\pi}\frac{\epsilon \rho_0H\Delta x\Delta y}{\bar{n}_pN_z},
\end{equation}
where $\bar{n}_p=1$ is the average number of particles per cell and $N_z$ is the number of cells in the vertical ($z$) direction.
Note that $\Delta y = 1$ for our 2D simulations.

We perform computations using the PPM algorithm with the Roe Riemann solver and the FARGO orbital advection scheme \citep{mig12} through which the boundary conditions in the radial (x) direction become simply periodic.
We employ $576^2$ grid zones in total (equivalent to $1440$ zones per scale height) and evolve the system up to $1000\,P$, where $P=2\pi/\Omega$ is the local orbital period.

Results showing the dust density distributions at different time are shown in Fig. \ref{fig:sb_maps}. The streaming instability is leading to dust clumping and concentrations. After 300 orbits we observe the start of larger clumps, which is often called the secondary phase of dust concentration and which was also reported in \citet{yan17}. 

In the top panel of Fig. \ref{fig:sb_Hp} we plot the particle scale height $H_p/H$ as a function of time with
\begin{equation}
  H_p = \sqrt{\overline{z_p^2} - \overline{z_p}^2}\,.
\end{equation}
The bottom panel of the same figure shows the maximum dust density as a function of time. Both results of the particles scale height and the dust concentration compare very well with the previous findings by \citep{yan17}.

}
\end{document}